\documentclass[aps,showpacs,groupedaddress,superscriptaddress,twocolumn,longbibliography]{revtex4-1}

\usepackage{float}
\usepackage{xcolor}
\usepackage{physics}

\usepackage{hyperref}
\hypersetup{
	breaklinks=true,
	colorlinks=true,
	allcolors=blue,
}

\usepackage{graphicx}
\usepackage{dcolumn}
\usepackage{bm}

\begin{document}
	
	\preprint{APS/123-QED}
	
	\title{Dynamics of quantum discommensurations in the Frenkel-Kontorova chain}
	\author{Oksana Chelpanova}
	\affiliation{ 
		Institut für Physik, Johannes Gutenberg-Universität Mainz, D-55099 Mainz, Deutschland}
	\author{Shane P. Kelly}
	\affiliation{Mani L. Bhaumik Institute for Theoretical Physics, Department of Physics and Astronomy, University of California at Los Angeles, Los Angeles, CA 90095}
	\author{Ferdinand Schmidt-Kaler}
	\affiliation{ 
		Institut für Physik, Johannes Gutenberg-Universität Mainz, D-55099 Mainz, Deutschland}
	\author{Giovanna Morigi}
	\affiliation{ 
		Theoretische Physik, Universität des Saarlandes, D-66123 Saarbrücken, Deutschland
	}
	\author{Jamir Marino}
	\affiliation{ 
		Institut für Physik, Johannes Gutenberg-Universität Mainz, D-55099 Mainz, Deutschland}
	\date{\today}
	
	\begin{abstract}
		
		The ability for real-time control of topological defects can open up prospects for dynamical manipulation of   macroscopic properties of solids. A sub-category of these defects, formed by particle dislocations, can be effectively described using the  Frenkel-Kontorova chain, which characterizes the dynamics of these particles in a periodic lattice potential. This model is known to host solitons, which are the topological defects of the system and are linked to structural transitions in the chain. This work addresses three key questions: Firstly, we investigate how imperfections present in concrete implementations of the model affect the properties of topological defects. Secondly, we explore how solitons can be injected after the rapid change in lattice potential or nucleated due to quantum fluctuations. Finally, we analyze the propagation and scattering of solitons, examining the role of quantum fluctuations and imperfections in influencing these processes. Furthermore,   we address the experimental implementation of the Frenkel-Kontorova model. Focusing on the trapped ion quantum simulator, we set the stage for controllable dynamics of topological excitations and their observation in this platform.
	\end{abstract}
	
	\maketitle
	
	\section{Introduction}

	In recent years, there has been a growing interest in the physics of incommensurate systems and the role of topological defects in altering the macroscopic properties of physical systems. This   has been particularly pronounced in the domain of nano-physics, where the commensurate-incommensurate (C-IC) transition is regarded as a   pinning-to-sliding transition occurring between two surfaces, making it a prominent item in the study of nano-friction ~\cite{Bonetti2021,hormann2023does,li2022nano,timm2021quantum,dong2023velocity,zheng2023structural,sar2023solvable,BRAUN200679,PhysRevResearch.3.043141}. 
	In materials science, the C-IC transition sparked curiosity in the late 1970s after experimental work by Naumovets and Fedorous~\cite{1977ZhETF..73.1085N}.  In their work, different interference peaks were reported  depending on the concentration of atoms absorbed by the surface of the substrate material (adatoms). A number of studies reported~\cite{AristovLuther,pokrovsky1984theory,bohlein2012observation} that when the concentration of the adatoms was small, they formed a commensurate structure with the surface, while for higher concentrations, the long-range Van der Waals force induced   an incommensurate configuration. The emergent lattice modulations have also been  shown to alter the properties of magnetic material. For instance, it has been shown that IC-C transition in the perovskite $R\text{MnO}_3$ is accompanied by the emergence of ferroelectric order at low temperatures~\cite{PhysRevLett.92.257201}. In general, the C-IC   transition has been shown to induce noteworthy modifications in the solid's macroscopic properties, piquing significant interest and setting the stage for future research~\cite{PhysRevLett.56.724,PhysRevB.54.17097,vanossi2013colloquium,woods2014commensurate,ahmed2023bicrystallography,Bak_1982,PhysRevB.66.073105,zhang2023spectroscopic,menu2023commensurateincommensurate}.

	In recent studies, the C-IC transitions  have been examined  from the perspective of its implementation in quantum simulators, including  driven Bose-Einstein condensates~\cite{Buchler_C_IC,KasperMarino,wybo2023preparing,mivehvar2021cavity,PolkovnikovQBreathers}, arrays of Rydberg atoms~\cite{Rydberg,PhysRevResearch.4.043102,chepiga2021kibble,zhang2024probing}, and trapped ion chains~\cite{Morigi3,Vuletic_nanofriction,vanossi2020structural,pyka2013topological}. What sets these quantum simulators apart is their unique capability for dynamic parameter control, allowing researchers not only to investigate the equilibrium C-IC transition but also to actively drive it in real time by adjusting the simulator's parameters. This approach offers the potential to dynamically alter the macroscopic properties of the system and manipulate its spectrum in real time. On the other hand, quantum fluctuations and finite-size effects, which are distinctive characteristics of quantum simulators, can pave the way for exotic phenomena that have not been observed on a macroscopic scale in traditional condensed matter systems~\cite{ejtemaee2013spontaneous,de2010spontaneous,ulm2013observation,solitons_short}.

	\bigskip

	In this paper, we study the  discommensurations  in such simulators, concentrating on the parameter regime accessible with the trapped ion simulator. We consider a one-dimensional chain containing up to a hundred charged particles within a periodic lattice potential captured by the Frenkel-Kontorova model (FKM)~\cite{frenkel1938see,FrankVanDerMerwe,braun2004frenkel}. This enables us to explore various effects and imperfections, such as discreetness and finite-size effects, as well as quantum fluctuations, and highlight the key distinctions between the C-IC transition in the quantum simulator and the C-IC transition in classical systems.

	Furthermore, we demonstrate how topological defects, holes, or double-filled sites (kinks and anti-kinks of the model) can be dynamically injected into the system by adjusting the amplitude of the lattice potential. Also, we show how quantum fluctuations can lead to nucleation of topological defect, resulting in a state that is a quantum superposition of commensurate and incommensurate chains.
	
	We delve further into the impact of simulator imperfections on the system's dynamics. To this end, we study the scattering of solitons in the model and show how finite-size effects, along with the quantum fluctuations, alter such scattering, comparing with the ideal scenario. 
	
	This paper is an extension of our previous work~\cite{solitons_short}, in which we solely studied the injection of the kink through an abrupt change in lattice periodicity.

	\subsection*{Summary of results}
	
	The paper is dedicated to exploring three main directions listed below.
	\begin{itemize}
		\item \textbf{Impact of    imperfections in the quantum simulation of solitons.}  In Sec.~\ref{sec:preliminaries}, we summarize the role of  discreteness, finite size, and quantum fluctuations  on soliton dynamics. In particular, we review the conditions under which the long-wavelength limit of the FKM is applicable. We {discuss features that break the integrability of the model, namely the corrections to the continuum limit due to the fact that the system is effectively discrete, which we will refer to as ``discreteness effects''. Moreover, we will discuss finite-size effects. We then} summarize different limits of the FKM  in Sec.~\ref{sec:FK_summary} and discuss possible implementation  of the dynamics we predict in a trapped ion simulator in Section~\ref{sec:exp}.
		
		\item \textbf{Real-time control of the platform and solitons injection.} 
		In Sections~\ref{sec:DynamicalResponse}, we show how solitons can be injected into the system from its boundaries in a controlled fashion. We consider driving dynamics by adjusting the amplitude of the lattice potential. Interestingly,  in this case, we also distinguish the generation of a bounded pair of kink and anti-kink, which is not observed when the lattice periodicity is instead varied, cf. Ref.~\cite{solitons_short}. 
		
		\item \textbf{Scattering of solitons as a test of integrability breaking in the FKM.} In Sec.~\ref{sec:Scattering}, we revisit kink-anti-kink scattering in the presence of quantum fluctuations and {accounting for} discreteness effects and {discuss} the difference from solitons scattering in the {continuum limit} of the model. Finally, we study a specific case of the scattering of the kink against the boundary of the Frenkel-Kontorova chain. We distinguish different outcomes of such scattering, namely, reflection and mirroring of the soliton or injection of multiple solitons. We provide explanations for these outcomes based on energy conservation arguments.
	\end{itemize}
	
	\begin{figure*}[]
		\includegraphics[width=0.99\linewidth]{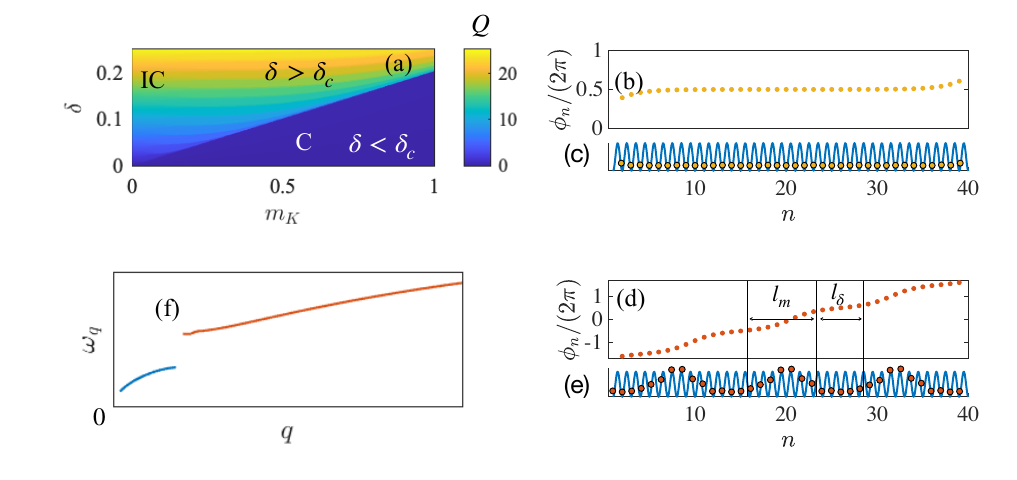}
		\caption{{Equilibrium properties of the FKM.} (a) The equilibrium phase diagram in terms of the lattice potential amplitude $m_K$ and the misfit parameter $\delta$. The number of solitons {$Q$ is encoded in the} colormap. (b) Phase $\phi_n$ for the commensurate configuration. (c) Sketch of particle positions on top of the lattice potential in the commensurate phase. (d) Phase $\phi_n$ for the incommensurate configuration. Each kink describes the local distribution of the empty lattice site. (e) Sketch of particle positions on top of the lattice potential in the incommensurate phase. (f) Sketch of the spectrum of the FKM in the incommensurate phase.}\label{fig:schematics}
	\end{figure*}

	\section{\label{sec:preliminaries}Preliminaries}
	This section is dedicated to giving a comprehensive introduction to the FKM. To highlight the main properties, we compare {the predictions of} this model with {the ones of} its continuous integrable counterpart. We discuss what breaks the integrability in the FKM and how such integrability breaking affects the dynamics of the system. Furthermore, we explore the influence of boundary conditions and finite-size effects, demonstrating how these elements can be leveraged to dynamically control topological defects within the system. Additionally, we briefly discuss an efficient approach for incorporating quantum effects into the dynamics through semi-classical methods. At the end of the section, we summarize  various parameter regimes that offer insights into the diverse physics encapsulated by the FKM. Readers already familiar with these concepts can proceed to the following section.
	
	\subsection{The Frenkel-Kontorova model}
	
	We consider a one-dimensional chain comprising  $N$ charged particles. {Due to the confinement, the particles are at uniform inter-particle spacing $a_0$. In addition, the particles also interact} with a substrate lattice potential with periodicity $a_s$. The displacement between the particle positions $x_n$ and the closest minimum of the external lattice, $ n a_s$,  is expressed in terms of a phase: 
	$\phi_n=2\pi(x_n-n a_s)/a_s-\pi$,
	where $n=1,\ldots, N.$ 
	The phase $\phi_n$ takes value $\pi$ when the ions' equilibrium positions coincide with a lattice minima. The dynamics of this phase are
	captured by the Frenkel-Kontorova model~\cite{frenkel1938see,FrankVanDerMerwe}
	\begin{equation}\label{eq:model}
		H=\sum_n \frac{p_n^2}{2} +{m_K^2}(1+\cos\phi_n)+\sum_{r=\pm 1} \frac{(\phi_{n+r}-\phi_n-2 \pi r \delta )^2}{2}
	\end{equation}
	where $p_n=\dot{\phi}_n$. 
	Here, the first term describes the system's kinetic energy, the second term models the external periodic lattice potential, and the last term accounts for the Coulomb repulsion between particles, {here truncated to the nearest-neighbor terms}. The parameter $m_K^2$ captures the amplitude of the periodic lattice potential, while the misfit parameter $\delta$  measures  the discommensuration of the system and is defined as a relative difference between two length scales of the model,  $\delta=(a_0-a_s)/a_s.$ 
	In  Eq.~\eqref{eq:model}, we set all variables and parameters to be dimensionless, {and, if necessary, rescaled in such a way that $|\delta|<0.5$}. A detailed derivation of the model~\eqref{eq:model} along with the rescaling can be found in Sec.~\ref{sec:exp}.
	
	The equations of motion that govern the dynamics of the finite-size FKM  read
	\begin{equation}\label{eq:eom}
		\begin{aligned}
			&	\ddot{\phi}_1={m_K^2}\sin \phi_1+\left(\phi_{2}-\phi_1 -{2\pi } \delta\right)\\
			&		\ddot{\phi}_{n=2:N-2}={m_K^2}\sin \phi_n+\left(\phi_{n+1}-2\phi_n +\phi_{n+1}\right)\\
			&		\ddot{\phi}_N={m_K^2}\sin \phi_N+\left(\phi_{N-1}-\phi_N +{2\pi } \delta\right).
		\end{aligned}
	\end{equation}
	Note that here, for particles in the bulk discommensuration effects from the left $+2\pi\delta$ and from the right $-2\pi\delta$ neighboring particles cancel each other, and thus, the misfit parameter does not directly affect the dynamics of this part of the chain. At the same time, the boundary sites are sensitive to the value of the misfit parameter $\delta.$
	
	By varying $\delta$, the system can undergo an equilibrium transition from the topologically trivial commensurate state to the incommensurate configuration  characterized by the presence of solitons, see  Fig.~\ref{fig:schematics}(a). When $\delta<\delta_c \approx 2 m_K / \pi^2,$  the external periodic potential dominates over inter-particle interaction, and particles are pinned by the substrate lattice, i.e., each particle sits in a minimum of the potential and $\phi_n=\pi.$
	Figure~\ref{fig:schematics}(b) displays the corresponding {phase} configuration, indicating that all particles are situated at the lattice minima positions, while Figure~\ref{fig:schematics}(c)
	illustrates the typical distribution of particles within the lattice in the commensurate state.

	In contrast, for $\delta>\delta_c,$  inter-particle interactions  destabilize the commensurate structure,  leading to particles rearrangement and the formation of dislocations, see Fig.~\ref{fig:schematics}(e). 
	These dislocations correspond to the local distribution of either holes ($\delta>0$) or excess particles ($\delta<0$) across the lattice and  can be mapped to the solitons  of the FKM, which are called respectively kinks or anti-kinks [see Fig.\ref{fig:schematics}(d)]. 
	Each kink (anti-kink) corresponds to the `step' in the phase $\phi_n,$ namely, the phase value changes by $2\pi$ ($-2\pi$) over a few lattice sites~\cite{Morigi3,Bak_1982,braun2004frenkel,Pruttivarasin_2011,garcia2007frenkel,Bak_1982}.
	The number of solitons
	\begin{equation} \label{eq:numkinks}
		Q=\frac{1}{2\pi}\left(\phi_N-\phi_1\right)
	\end{equation}
	is a quantity that allows us to distinguish between the commensurate ($Q=0$) and the incommensurate ($Q\ne 0$) phases~\cite{AristovLuther} by counting the total phase accumulated in units of $2\pi$. For instance, the incommensurate configuration in Fig.~\ref{fig:schematics}(d,e) has three `steps' in $\phi_n$  and corresponds to $Q=3.$  In the incommensurate phase, $Q$ is controlled by $\delta.$

	When the length of the kink is significantly larger than the lattice spacing, the {discrete charge distribution can be approximated by a continuum. The continuum limit of the FKM is the so-called}  Pokrovsky and Talapov  {model}\cite{pokrovsky1984theory}. Importantly, the {Pokrovsky and Talapov model} is integrable, and its analytical solutions are often employed  as a {benchmark for the FKM}. In the following section, we summarize predictions of the Pokrovsky-Talapov model {that are relevant to this study}. 
	
	\subsection{Long-wavelength limit of the Frenkel-Kontorova model}
	
	When the size of the discommensuration region is much larger than inter-particle spacing, the {phase function} $\phi_n$ can be replaced with the field $\phi(x)$, and the finite differences in~\eqref{eq:model} can be replaced by partial derivatives in space. The effective model can be then described by  the Pokrovsky-Talapov  Hamiltonian, which reads  $H=\int \mathrm{d}x h,$ where the Hamiltonian density is given by
	\begin{equation}\label{eq:PT}
		h=\frac{1}{2}p^{2}+\frac{1}{2}\left(\partial_{x}\phi\right)^{2}+\frac{m_K^{2}}{\beta^{2}}(1+\cos\beta\phi)-\frac{\mathcal{H}\beta}{2\pi}\partial_{x}\phi,
	\end{equation}
	where $p=\dot{\phi}$, $\beta$ calls the interaction constant, and parameter $\mathcal{H}$ sets the density of topological defects. The interplay between kinetic terms and non-linearities in the first three terms enables soliton solutions, while the last term  controls the density of these solitons within the chain. 
	
	The evolution of the field $\phi(x)$ according to Eq.~\eqref{eq:PT} follows the sine-Gordon equation~\cite{malard2013sine}
	\begin{equation}\label{eq:PT_EoM}	\partial_{t}^{2}\phi=\partial_{x}^{2}\phi+\frac{m_K^{2}}{\beta}\sin\beta\phi.
	\end{equation}
	The possible quasi-particle solutions of the model~\eqref{eq:PT_EoM} are determined  by the interaction constant  $\beta$~\cite{ruelle2009review,del2023exact}. From mapping between models~\eqref{eq:model} and \eqref{eq:PT} one can find $\beta=1.$ This specific choice enables kink (anti-kink) solutions. Additionally, the model can host a high-energy breather solution, which is the bounded kink-anti-kink pair that oscillates back and forth around a common center.

	The important feature of the chain governed by the Pokrovsky-Talapov model is that the density of the defects in it can be controlled by adjusting the parameter $\mathcal{H}$, cf. Refs.~\cite{AristovLuther,KasperMarino}. The last term in Eq.~\eqref{eq:PT}   bears the analogy with the chemical potential in statistical physics~\cite{huang2009introduction}. However, instead of controlling the number of particles in a macro-canonical ensemble, it  controls the number of topological defects in the chain.
	
	In particular, the so-called `chemical potential' $\mathcal{H}$ sets the boundary conditions where the equilibrium solution seeks to minimize the total energy, cf. Refs.~\cite{AristovLuther,KasperMarino,Lazarides_finite_T_RG},
	\begin{equation}
		E=\int_0^L dx\left(\frac{1}{2}\left(\partial_x \phi\right)^2+\frac{m_K^2}{\beta^2}(1+\cos \beta \phi)\right)-\mathcal{H}Q_c.
	\end{equation}
	Here, we denote $Q_c=\beta/(2\pi)(\phi(L)-\phi(0))$. This parameter quantifies the change in the field $\phi$ over the length of the chain, counting the number of kinks present within it. Adjusting the `chemical potential' $\mathcal{H}$ makes it  possible to steer the system across various meta-stable minima within the energy landscape. Each of these minima corresponds to distinct soliton configurations within the chain~\cite{Lazarides_finite_T_RG}.
	
	It has been shown~\cite{AristovLuther,KasperMarino} that in equilibrium, the  ground state of  the Pokrovsky-Talapov model reads
	\begin{equation}\label{eq:kink}
		\phi(x)=\left\{
		\begin{aligned}
			&\pi/\beta,&\quad \mathcal{H} < M_s,\\
			&\frac{2}{\beta}\operatorname{am}\left(\dfrac{x m_K}{k},k^2\right),&\quad{\mathcal{H}\ge M_s}.
		\end{aligned}
		\right.
	\end{equation}
	\noindent Here, the parameter $k$ depends on $m_K$ and $\mathcal{H}$, $M_s=8 m_K/\beta^2$ is the soliton mass, and $\operatorname{am}()$ denotes the Jacobi amplitude function~\cite{abramowitz1988handbook}. If $\mathcal{H}<M_s$, the energy of the incommensurate configuration exceeds that of the commensurate chain. In this case, in the ground state,  the particles organize themselves in the equidistant configuration with $\phi(x)=\pi/\beta$. On the other hand, when $\mathcal{H}\ge M_s$, the ground state {contains a}  finite density of discommensurations. In the ground state, the field adopts a staircase configuration, where each step has a height $2\pi/\beta$ and corresponds to a kink. The length of a single kink, $l_{m}\propto 1/m_K$, represents the size of the region in which $\phi$ changes by $2\pi/\beta$. Conversely, the distance between kinks is controlled by the chemical potential $l_{\delta}\propto 1/\mathcal{H}$. Note that these results qualitatively describe the corresponding commensurate-incommensurate transition in the FKM mentioned above, where the correspondence between chemical potential and the misfit parameter is given by $\mathcal{H}=4\pi^2 \delta/\beta^2$.
	
	\bigskip
	
	Depending on the resulting ground state, the model~\eqref{eq:PT} can host various types of excitations. The excitation spectrum of the model~\eqref{eq:PT} can be derived by expanding field $\phi(x) $ on top of the ground state configuration $\phi^{(0)}(x)$ in powers of fluctuations $\eta(x)$ as $\phi(x)=\phi^{(0)}(x)+\eta(x).$  In the lowest order, which is valid in the low energy limit,  this expansion leads to the   quadratic Hamiltonian~\cite{AristovLuther} 
	\begin{equation}\label{eq:ex_spectrum}
		\mathcal{H}_{\eta}=\frac{1}{2}\left(\partial_t \eta\right)^2+\frac{1}{2}\left(\partial_x \eta\right)^2-\frac{1}{2} m_K^2\left(\cos \beta \phi^{(0)}\right) \eta^2+\ldots
	\end{equation}
	The equations of motion governing the dynamics of the excitations read
	\begin{equation}\label{eq:spectrum}
		\left(-\partial_x^2-m_K^2 \cos \beta \phi^{(0)}\right) \eta_i(x)=\omega_i^2 \eta_i(x).
	\end{equation}
	Here, the index $i$ {labels} the $i$-th mode. {In} the commensurate phase, the spectrum exhibits a gap proportional to $m_K$. Following the terminology in Ref.~\cite{AristovLuther}, which is inherited from the solid state physics~\cite{ashcroft2001festkorperphysik}, we refer to this branch of the spectrum as the optical modes. The optical modes govern individual excitations of particles around their equilibrium positions, with frequencies proportional to $m_K$. 
	
	In the incommensurate phase, the spectrum contains two branches.  The first branch is formed from the hybridization of the bound states related to individual $Q_c$ kinks in the chain. These $Q_c$ modes correspond to gapless acoustic excitations that control  the propagation of $Q_c$ kinks along the chain~\cite{AristovLuther,PhysRevE.52.3892}. The second branch describes individual oscillations of the rest of the particles around their equilibrium positions.

	\bigskip
	
	The integrable Pokrovsky-Talapov model~\eqref{eq:PT} provides valuable insights and foundation for developing an intuitive understanding of the commensurate-incommensurate transition and the   characteristics of solitons. Nevertheless, in realistic implementations, numerous factors can disrupt integrability and significantly alter the chain's behavior. In the following subsections, we summarize the impact of finite-size effects, the chain's discreteness, and quantum corrections on the  system's dynamics. We also establish the conditions under which these effects are negligible, and the continuous description remains valid.

	\subsection{ Discreteness effects\label{sec:PNpotential}}
	
	In this section, we summarize the role of  discreteness effects and determine a parameter regime within which the continuum model~\eqref{eq:PT} can provide accurate predictions for   soliton dynamics. The FKM is a discrete lattice model, with the amount of `discreteness' controlled by the amplitude of the lattice potential, $m_K.$ Depending on $m_K$, the dynamics (propagation) of the kinks may {exhibit very different features}: In the limit $m_K\to 0$, solitons can propagate ballistically through the lattice, resembling the behavior commonly observed in integrable models. On the other hand, it has been reported  that in the {regime where the continuum approximation becomes invalid} ($m_K\propto 1$), solitons do not propagate `frictionlessly' through the lattice.  Instead, they slow down during dynamics or even become immovable after short periods of time~\cite{braun2004frenkel,peyrard1984kink,askari2020collision,PhysRevB.33.1904}. As we show in Sec.~\ref{sec:KAKScattering}, the discreteness effects become important in scattering processes since kinematic characteristics of solitons before collision are crucial for the outcome of the scattering. 
	
	\bigskip
	
	The role of discreteness effects {on the continuous model} can be understood from a symmetry point of view: In the continuous model, the system possesses continuous translation invariance, which is broken in the incommensurate phase and restored in the commensurate phase. The breaking of this symmetry results in the appearance of the gapless Goldstone mode in the excitation spectrum, the acoustic mode mentioned earlier, which governs the propagation of kinks through the lattice~\cite{AristovLuther,KasperMarino}. In the discrete model, {instead}, continuous translation invariance is reduced to a discrete one, and its breaking in the incommensurate phase is associated with the appearance of  a gaped Peierls-Nabarro mode in the spectrum~\cite{braun2004frenkel}, cf. Fig.~\ref{fig:schematics}(f). The difference in the nature of the acoustic mode in continuous and discrete regimes leads to distinct dynamical properties.
	
	We can map the discrete model to the continuum one by substituting $\phi_n\to\phi(x)$. For the finite differences in Eqs.~\eqref{eq:eom}, we can employ a Taylor series expansion, which takes the form
	\begin{equation}
		\phi_{n+1}-2\phi_n+\phi_{n-1}\to \frac{\partial ^2\phi}{\partial x^2}+
		\frac{2}{4!}\frac{\partial ^4\phi}{\partial x^4}+\ldots.
	\end{equation}
	The sine-Gordon model is integrable, while including higher-order derivatives disrupts this integrability and significantly impacts the model's properties.  However,  these higher-order corrections {become negligible by decreasing} the amplitude of the potential $m_K.$ Qualitatively, this stems from the fact that in the $m_K\to 0$ limit the length of the kink  $l_m\propto 1/m_K\to\infty$, and the kink appears wide and smooth. Thus, higher-order derivatives $\partial^{2n}\phi/\partial x^{2n}$ are suppressed compared to the contribution from  the $\partial^{2}\phi/\partial x^{2}$ term. As $m_K$ increases, the influence of higher-order terms becomes more significant in shaping the dynamics; thereby, the model~\eqref{eq:PT} does not approximate FKM correctly. 
	
	\bigskip
	
\begin{figure}[]
		\includegraphics[width=0.99\linewidth]{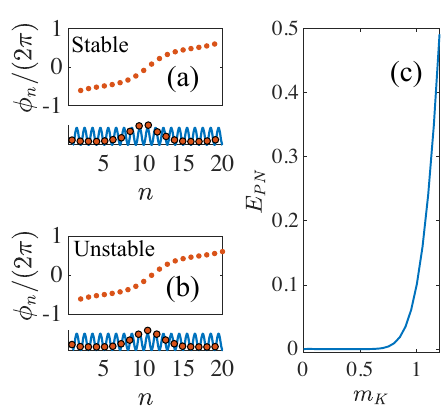}
		\caption{{The Peierls-Nabarro barrier.} 
			(a-b) Two stationary kink configurations (compare the particles' positions in the middle of the chain). The Peierls-Nabarro barrier is determined as the difference in the potential energies of configurations (b) and (a).
			(c) The Peierls-Nabarro barrier as a function of $m_K.$  
		}\label{fig:PN}
	\end{figure}

	To estimate the value of $m_K$ at which the continuous model~\eqref{eq:PT} fails to accurately predict the kinematics of a single kink {of the FKM}, one can consider the propagation of a kink across a single lattice site. During this process, the positions of the particles neighboring the kink change, resulting in the change of  the system’s potential energy along the path, $V_{\operatorname{pot}}=\sum_n m_K^2(1+\cos \phi_n)+\sum_{r= \pm 1} (\phi_{n+r}-\phi_n-2 \pi r \delta)^2/{2}$. Thus, to propagate to the following site, the kink should overcome a Peierls-Nabarro barrier $E_{PN}$~\cite{braun2004frenkel,ablowitz2021peierls,partner2013dynamics,PhysRevB.33.1904}, which is the difference between the maximal, see Fig.~\ref{fig:PN}(b), and minimal, see Fig.~\ref{fig:PN}(a), potential energy of the system through this process
	\begin{equation}~\label{eq:EPN}
		E_{PN}=V_{\operatorname{pot}}(\phi_{\operatorname{kink}}^{\operatorname{unstable}})-V_{\operatorname{pot}}(\phi_{\operatorname{kink}}^{\operatorname{stable}}).
	\end{equation}
	The barrier $E_{PN}$ has a clear physical meaning as the minimum kinetic energy the kink must have to move to the following site. If the kink cannot overcome this barrier, it  gets stuck at the initial site. When the kink propagates through the lattice, it needs to climb the $E_{PN}$ barrier, which results in a decrease in its kinetic energy. The released kinetic energy of the kink is transferred to the lattice vibrations in the form of the optical modes, and  surrounding particles start oscillating  around  their equilibrium positions. 
	
	The dependence of the Peierls-Nabarro barrier on $m_K$ is plotted in Fig.~\ref{fig:PN}(c). One can recognize that for small $m_K$, the kink can freely propagate through the lattice, while for $m_K\gtrsim 0.8$, the Peierls-Nabarro barrier increases. For $m_K>1.1$, a single kink can propagate only a few lattice sites before it comes to a complete stop.

	\subsection{Finite-size and boundary effects}
	
	The finite-size and boundary effects play a crucial role in the ability to dynamically  manipulate topological defects. The finite system size makes it necessary to treat exactly the boundary condition of the model, which, as opposed to the continuous model, brings a direct dependence of the dynamics of the system on the misfit parameter $\delta.$

	In the FKM, the boundaries act as a source of topological defects with some chemical potential, and thus, the corresponding boundary conditions are different from the ones often considered in the continuum  Pokrovsky-Talapov model~\cite{AristovLuther}. Precisely,  equations of motion~\eqref{eq:eom} are directly sensitive to the misfit parameter through the boundary sites, enabling the possibility to dynamically pump kinks or anti-kinks into the system from the boundaries, thereby altering the macroscopic properties of the chain, as we illustrate in  Sec.~\ref{sec:DynamicalResponse}.

	Additionally, the boundary conditions of the  finite-size FKM [cf. Eqs.~\eqref{eq:eom}] result in the dressing (i.e., minor modification) of the number of kinks $Q$.  For small but non-zero values of the misfit parameter, $0<\delta<\delta_c$, the equilibrium value of the phase $\phi_{1,N}$ deviates from $\pi$ by the quantity, linear in $\delta.$   Due to this fact, the number of kinks $Q$ [cf. Eq.~\eqref{eq:numkinks}] also deviates  from integer by the value linear in $\delta $ even in the commensurate phase. This additional soliton density is induced solely by the boundary conditions scales like $\delta/N$ with the system size, and it becomes irrelevant only for an infinite system. However, such boundary defects in finite-size chains can have a pronounced effect on scattering processes, see Sec.~\ref{sec:scatteringonboundary}.
	
	\bigskip
	
	A second finite size effect comprises modulation of the soliton density with the system size. Following the Pokrovsky-Talapov model, the total length of one kink equals $l_m+l_{\delta}.$ Thus, even if the system size is varied within $[N,N+l_m+l_{\delta})$ in equilibrium, the number of kinks remains constant, determined by the number of kinks that can fit a given chain, namely $Q= \lfloor N/(l_m+l_{\delta})\rfloor$. The injection of another kink requires additional energy costs due to the effective repulsion between kinks. As a result, the kinks density, which is equal to $\rho=Q/N$ varies periodically with the system size in the range $\rho=(Q/N,Q/(N+l_m+l_{\delta})),$ see Ref.~\cite{KasperMarino} for comprehensive discussion. 
	
	\bigskip
	
	Finally, if the system size is smaller than $l_m+l_{\delta}$, the form of a kink will be modified to make it sharper and fit the lattice. This change of the kink's shape requires additional energy cost, resulting in a higher equilibrium critical value of the misfit parameter $\delta_c$ than predicted by the Pokrovsky-Talapov model. In this regime, a field theory~\eqref{eq:PT} fails to predict system properties. Our analysis, however, is in the regime where at least a few solitons fit into the system. 
	
	\subsection{Quantum mechanics of the Frenkel-Kontorova model\label{sec:preliminaries_TWA}}
	
	The quantum simulators of the FKM operate on length and energy scales where quantum fluctuations can play a significant role and, therefore, should be taken into account. Quantum fluctuations may destroy ordered phases, generate new dynamical phases, or give rise to a quantum critical region where the transition from one phase to another appears as a crossover \cite{sachdev,vestigial_order,Giovanna_zigzag_PRL, nucleation,Giovanna_zigzag_PRA,Lazarides_finite_T_RG,AlessioPRL,AlessioPRB,marino2022dynamical,BerdanierBoundary,wybo2023preparing}. Consequently, we  extend the model~\eqref{eq:model}  such that is
	\begin{equation}\label{eq:qm_model}		\hat{H}=\sum_n\frac{\hat{p}_n^2}{2} +m_K^2(1+\cos\hat{\phi}_n)+\sum_{r=\pm 1} \frac{(\hat{\phi}_{n+r}-\hat{\phi}_n-2 \pi r \delta )^2}{2}.
	\end{equation}
	Here we  introduce momentum and coordinate operators $\hat{p}_n,$ $\hat{\phi}_n$ instead of classical variables $p_n,\phi_n.$ These operators satisfy canonical commutation relations 
	\begin{equation}  \left[\hat{\phi}_n,\hat{p}_m\right]=ih_{\operatorname{eff}}\delta_{n,m}.
	\end{equation}
	Here, the effective Plank constant 
	is a dimensionless small parameter that controls the strength of the quantum fluctuations. The value of $h_{\operatorname{eff}}$ depends on the particular experimental implementation of the model~\eqref{eq:model}. Details of Eq.~\eqref{eq:qm_model} and $h_{\operatorname{eff}}$ are found in  Sec.~\ref{sec:exp}. 
	
	In the limit $h_{\operatorname{eff}}\to 0$, quantum fluctuations can be considered as a minor correction to the classical effects and thus treated perturbatively. In this regard, one can build a Bogoliubov-Born-Green-Kirkwood-Yvon (BBGKY) hierarchy of corrections in powers of $h_{\operatorname{eff}},$ when higher-order corrections contribute to dynamics at later timescales~\cite{negele2018quantum}. 
	
	In the following, we  treat quantum corrections in a semi-classical limit, where only corrections linear in $h_{\operatorname{eff}}$ are included in the dynamics. To do so, we follow the truncated Wigner approximation (TWA) and consider $\hat{\phi}_n$ and $\hat{p}_n$ as a sum of the classical term and the small quantum correction~\cite{POLKOVNIKOVTWA,TWA1stcorr,PNAS_TWA}.  For simplicity, we initialize our system in the commensurate phase, for which the ground state can be approximated by the Gaussian wave function with the Wigner quasi-probability distribution 
	\begin{equation}\label{eq:TWA}
		W\left(\eta_{q}^{0},n_{q}^{0}\right)=\prod_{q}\exp\left[-{\left|\eta_{q}^{0}\right|^{2}}/{\sigma_{q}}-{\left|\psi_{q}^{0}\right|^{2}}/b_{q}\right],
	\end{equation}
	\noindent where $\eta^0_q=\phi^0_q-\langle \hat{\phi}_q\rangle_0$ and $\psi^0_q=p^0_q-\langle \hat{p}_q\rangle_0$ are the phase space variables relative to the mean value of the operators in the initial state,  while $\sigma_q=\langle \hat{\eta}_q^2\rangle_0=h_{\operatorname{eff}}/\omega_q$ and $b_q=\langle \hat{\psi}_q^2\rangle_0=h_{\operatorname{eff}}\omega_{q}$ encode the variance  of quantum fluctuations in the initial state. Here, $\omega_q$ is determined by the discrete counterpart of the Eq.~\eqref{eq:spectrum}, cf. Fig.~\ref{fig:schematics}(f).  At zeroth order in $h_{\operatorname{eff}}$, the TWA approximation is simply the mean-field dynamics as in Eq.~\eqref{eq:eom}.  At first order in $h_{\operatorname{eff}}$, the dynamics of phase space variables are still described by Eq.~\ref{eq:eom}, but the initial state is sampled from the positive Wigner distribution in Eq.~\eqref{eq:TWA}. Observables at later times are calculated by averaging over the trajectories resulting from the initial quantum noise. Taking into account higher-order corrections implies adding quantum noise during dynamics. Precisely, at randomly distributed time moments, the momentum $p_m$ undergoes additional jumps induced by quantum fluctuations~\cite{POLKOVNIKOVTWA,TWA1stcorr}. 
	
	\subsection{Summary of regimes of the Frenkel-Kontorova model\label{sec:FK_summary}}
	
	\begin{figure}
		\includegraphics[width=0.99\linewidth]{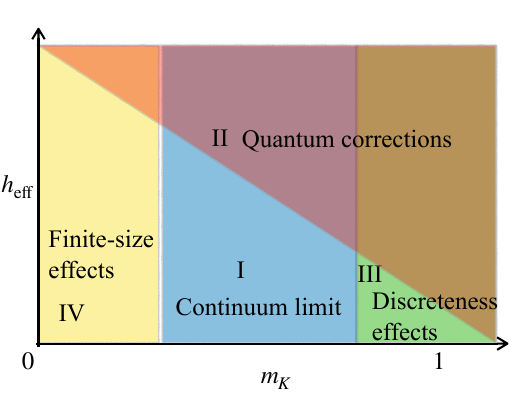}
		\caption{Cartoon of possible limits of the FKM depending on the $m_K$  and the strength of $h_{\operatorname{eff}}.$
			(I) Range of values of $m_K$ in which the Pokrovsky-Talapov model governs dynamics correctly. Note that for higher values of $m_K$ quantum corrections are  required to capture dynamics. (II) Region of parameters in which quantum fluctuations should be taken into account to recover the dynamics of the system correctly. (III) For $m_K\gtrsim 0.8 $, discreteness effects provoke significant dissipation of the soliton's kinetic energy. In this limit, the Pokrovsky-Talapov model is invalid. (IV)	When $1/m_K \gtrsim N$, where $N$ is the system size, the length of a single kink exceeds the system size, resulting in a change in the critical misfit parameter compared to   the Pokrovsky-Talapov model predictions.
		}\label{fig:cartoon}
	\end{figure}
	
	The sine-Gordon model is integrable in classical  and  quantum limits~\cite{del2023exact}. Thus, the sources of the integrability breaking in the model are discreteness and finite size effects but not a finite $h_{\operatorname{eff}}$ value. As a result, one can probe different regimes, integrable or non-integrable, classical or quantum, by varying $m_K$ and $h_{\operatorname{eff}}$ separately.

	The illustrative sketch in Fig.~\ref{fig:cartoon} provides a concise summary of the diverse regimes within the FKM that have been discussed thus far. When $m_K\to 0,$ one can obtain a regime in which continuous description works relatively well, and solitons behave like solitons of the integrable sine-Gordon model (I). When $m_K$ increases and Peierls-Nabarro barrier $E_{PN}$ becomes finite (cf. Sec.~\ref{sec:PNpotential}), solitons start experiencing dissipation of their kinetic energy to the lattice vibrations (III). Along with that, by adjusting the value of the effective Plank constant, one can make the effect of quantum fluctuations sizable. As elaborated in subsequent sections, these fluctuations significantly influence the critical region, giving rise to a novel dynamical regime characterized by a superposition of chains with and without solitons (II).

	Furthermore, another parameter range can be explored when dealing with an exceptionally weak lattice potential amplitude, $m_K\to 0$, where the length of a single kink surpasses the system's dimensions (IV). In such instances, the critical misfit parameter at which the transition occurs becomes sensitive to the system's size and deviates from the predictions of the Pokrovsky-Talapov model. 
	
	In the distinct parameter regimes depicted in Fig.~\ref{fig:cartoon}, the impact of finite size, discreteness, and quantum effects extends beyond the equilibrium properties of the system to also influence its dynamics. The following section  provides a more detailed exploration of these effects, focusing on the system's dynamical response to abrupt parameter changes.

	\section{\label{sec:DynamicalResponse} Real-time control of the platform and solitons injection}
	
	\begin{figure}[]
		\centering
		\includegraphics[width=0.75\linewidth]{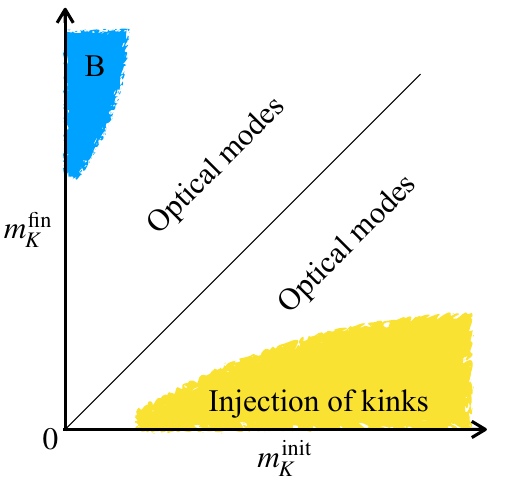}
		\caption{ Sketch of the dynamical phase diagram in terms of the initial $m_K^{\operatorname{init}}$ and final $m_K^{\operatorname{fin}}$ values of $m_K$. For some combination of parameters,  only the optical mode can be excited (white central region), solitons can be injected (yellow), or the breather can be excited (blue).
		}
		\label{fig:mk}
	\end{figure}
	
	In this section, we explore structural C-IC  transitions dynamically driven by quenches of the lattice amplitude $m_K$. We start with the classical treatment and discuss different modes that can be excited by varying $m_K$. We show how, in this way, one can inject topological defects into the system, altering the system's properties.  Then, we explore how quantum fluctuations modify the dynamics of the systems in proximity to the boundary that separates the commensurate and incommensurate phases. Finally, we provide insights into the distinctions between fast quenches and the gradual, adiabatic adjustment of parameters.
	
	\subsection{Excitations}
	
	We initialize the system in the equilibrium configuration with given initial $m_K^{\operatorname{init}},$ and then evolve it with some final value $m_K^{\text{fin}}$. Throughout this process, we maintain a constant misfit parameter $\delta$. Various dynamical responses on such quench are summarized in Fig.~\ref{fig:mk}. Here, we find three possible outcomes of the quenching  $m_K$ in the FKM
	\begin{enumerate}
		\item The number of kinks remains constant, while small excitations of the particles around their equilibrium positions are excited as an optical mode; this corresponds to the white region in Fig.~\ref{fig:mk};
		\item Injection of kinks from the boundaries of the chain, accompanied via excitation of the acoustic modes; this corresponds to the yellow region;
		\item Excitation of a bounded superposition of the kink and anti-kink (breather); this corresponds to the blue region.
	\end{enumerate}
	We explore these regimes in more detail below.

	\subsubsection{Excitation of the optical modes after the quench of $m_K$} 
	
	\begin{figure}[]
		\includegraphics[width=0.99\linewidth]{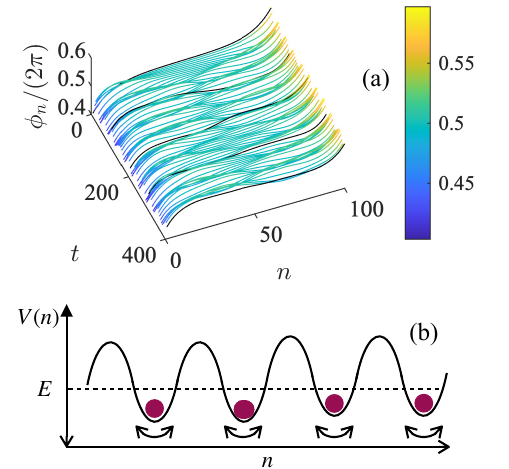}
		\caption{(a) Time evolution of the $\phi_n$ after the quench of $m_K.$ For small changes of $m_K$, particles start  oscillating around their new equilibrium positions. (b) Sketch of the particle dynamics in the chain. Here $m_K^{\operatorname{init}}=0.1$, $m_K^{\operatorname{fin}}=0.15$, and $\delta=0.01$.
		}\label{fig:optical_mode}
	\end{figure}
	
	Referring to the equilibrium phase diagram in Figure~\ref{fig:schematics}(a), it is evident that the parameter $m_K$ can be adjusted without inducing a transition between commensurate and incommensurate phases, namely without change of the soliton's number $Q$.   In real-time dynamics, such quench of $m_K$ can be seen as a tiny perturbation since the energy injected with the quench remains below the threshold required to inject kink or breather excitations. In this process, the energy difference between two stable configurations for initial $m_K^{\operatorname{init}}$ and final $m_K^{\text{fin}}$  is transferred into vibration modes, see the white region in Fig.~\ref{fig:mk}.

	The quenching of $m_K$ is achieved through a rapid change of the lattice amplitude, and its effects can be comprehended by examining equilibrium solutions for both the initial and final values of $m_K.$ In the incommensurate phase, $m_K$ sets the length of a single kink, with $l_m$ being inversely proportional to $m_K$ (i.e., $l_m\propto 1/m_K$). Consequently, when the lattice amplitude is adjusted, the equilibrium spacing between kinks also changes. Particles that are initially positioned at their equilibrium locations before the quench are no longer in equilibrium afterward, as depicted in Fig.~\ref{fig:optical_mode}(b). These particles, displaced from their equilibrium positions, commence oscillating around new equilibrium positions determined by $m_K^{\text{fin}}.$ The difference in energy between the state after the quench and the equilibrium configuration with $m_K^{\operatorname{fin}}$ contributes to the excitation of optical modes. The frequency of these oscillations is determined by the $m_K^{\text{fin}}$, and the amplitude of the oscillations is set by the amount of the exceeding energy.
	
	\bigskip
	
	For quenches within the commensurate phase, the displacement of  particles in the bulk in equilibrium  for both values of $m_K^{\operatorname{init}}$ and $m_K^{\operatorname{fin}}$ read $\phi_n=\pi$.  Thus, after the quench, these particles are not displaced from equilibrium and remain at rest. However, starting from the commensurate phase, the optical mode can still  be excited after the quench of $m_K$. This happens because, for any finite misfit parameter $\delta\ne 0$, the deviation of the phase at the boundaries is non-zero and depends on $\delta$ and $m_K$. Thus, after the quench of $m_K$ optical mode is excited at the boundary sites, and then this excitation propagates due to the nearest neighbor interaction into the bulk with the speed of sound $c=1$, see Fig.~\ref{fig:optical_mode}(a).
	
	If $m_K$ is varied slightly, only optical modes can be excited in the system, cf. Fig.~\ref{fig:mk}. As we will show below, the injection of a soliton and the excitation of a breather require higher energy and more substantial system perturbation.

	\subsubsection{Injection of kinks accompanied via excitation of the acoustic mode\label{sec:sonic}}
	
	\begin{figure}
		\includegraphics[width=0.99\linewidth]{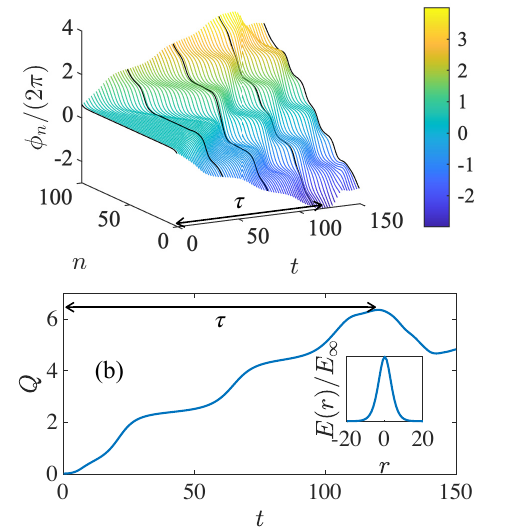}
		\caption{(a) Evolution of $\phi_n$ after the quench of $m_K^{\operatorname{init}}\to m_K^{\operatorname{fin}}\le m_K^c$ when kinks periodically enter the system from both boundaries. (b) Number of kinks in the system as a function of time. The inset shows the energy of two kinks in equilibrium as a function of the distance between their centers. Here $m_K^{\operatorname{init}}=0.5$, $m_K^{\operatorname{fin}}=0.25$, and $\delta=0.08$.}\label{fig:sonic1}
	\end{figure}
	
	According to the equilibrium phase diagram in Fig.~\ref{fig:schematics}, the commensurate-incommensurate transition can occur when keeping the misfit parameter fixed and decreasing $m_K.$ If we decrease the mass of the kink $m_K$ below a certain threshold, $m_K^{\operatorname{init}}\to m_K^{\operatorname{fin}}\le m_K^c$, the initial energy of the system can be sufficient to inject a finite density of kinks into the system. In this case, we lower the amplitude of the lattice potential, causing particles to redistribute in the chain and form local discommensurations, see the yellow region in Fig.~\ref{fig:mk}.

	The example of dynamics with  injection of solitons is illustrated in Fig.~\ref{fig:sonic1}(a). Starting from the commensurate configuration, kinks enter the chain one by one via the system boundaries. After a timescale of $\tau\propto N$, the density of kinks saturates, and then kinks traverse the lattice back and forth, scattering off each other. Importantly,  since the topological charge within the chain must be conserved, discommensurations can enter or leave the system solely at the boundaries, which can be considered as a reservoir of topological defects, see also Refs.~\cite{PhysRevA.91.033607,PhysRevLett.110.025302,PhysRevE.100.062202,ma2020realization,etde_6020750,dauxois2006physics,Brox_kinks_spectroscopy}.

	The number of kinks $Q$ in the system as a function of time is plotted in Fig.~\ref{fig:sonic1}(b). Just after the quench, $Q$ starts growing in a staircase fashion, and then it saturates around a certain value. One can recognize that the height of each `step' in the staircase is equal to two, as the injection of kinks occurs independently from both the left and right boundaries. The duration of each  plateau (a time interval during which the number of kinks $Q$ remains constant) is proportional to $l_{\delta}/c$, while the time intervals during which the charge changes from $Q$ to $Q+2$ are set approximately by $l_{m}/c$.

	The staircase structure of $Q(t)$ can be elucidated by analyzing the energies associated with various equilibrium configurations of the kinks. The inset in Fig.~\ref{fig:sonic1}(b) shows the energy of two kinks as a function of the relative distance, $r$, between their centers. If the kinks overlap, their combined energy is higher than the energy of the configuration with two infinitely separated kinks, resulting in effective short-distance repulsion~\cite{CARRETEROGONZALEZ2022106123,braun2004frenkel}. As such, the average distance between kinks after the quench is set approximately by $l_m+l_{\delta}$ because arranging kinks closer requires additional energy. Consequently, over longer times, the number of kinks tends to saturate at approximately $Q \approx 2N / (l_m + l_\delta)$. This quantity is primarily determined by the total number of solitons that can be accommodated within a system of a given size $N$. The factor of two arises from the behavior of the nearest neighbor interaction model, where the left boundary remains unaware of the injection of kinks through the right boundary until a time of $N / c$ when the first injected kinks reach the opposite boundaries. Accordingly, $N / (l_m + l_\delta)$ kinks are injected from the left boundary and an equal number from the right boundary, independently. This factor two disappears when particles interact with each other in a long-range fashion. In this case, the injection of kinks stops once the number of kinks reaches the value $N/(l_{\delta}+l_m),$ cf. Ref.~\cite{solitons_short}
	
	Until a time scale of  $\tau\propto N/c$, systems with different numbers of particles in the short-range interacting model evolve with the same staircase temporal profile, as their dynamics are universally governed by ${\delta}$ and $m_K$. However, in the long-range interacting model, the temporal profiles of $Q$ can vary depending on the system sizes. This mechanism is identical to the one reported in Ref.~\cite{solitons_short} for the case when a  quench in $\delta$ is performed.
	
	\bigskip
	
	Note that the critical value of $m_K^{cd}$ at which transition to incommensurate phase occurs in dynamics differs from the critical $m_K^c$ in equilibrium:  While in equilibrium, kinks are immovable, the dynamical phase transition is associated with the excitation of the acoustic mode, which governs the propagation of kinks. As such, the dynamical phase transition requires higher energy and happens at the smaller value of $m_K.$ We omit the apex $^d$ when discussing dynamical critical values for simplicity in the rest of the text but recover it again in Sec.~\ref{sec:Conclusions}.

	\begin{figure}
		\includegraphics[width=0.99\linewidth]{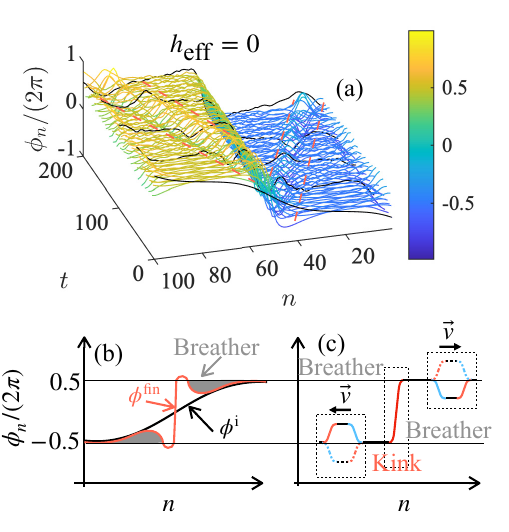}
		\caption{(a) The time evolution of $\phi_n$ subsequent to the quench, where  $m_K^{\operatorname{init}}$ is significantly less than $m_K^{\operatorname{fin}}$. Dashed red lines indicate the positions of the breathers. Here $m_K^{\operatorname{init}}=0.1$, $m_K^{\operatorname{fin}}=0.7$, and $\delta=0.025$. (b) The shape of a single kink before (black line) and just after (blue line) the quench. The gray shading is employed to denote discommensurations corresponding to the formation of breathers. (c) Sketch of $\phi_n$ near the center of the chain. After the quench of $m_K$, two breathers arise in the system. Each kink-anti-kink pair is characterized by a non-zero velocity $\pm {v}$, and the breathers move away from the kink. }\label{fig:breathers}
	\end{figure}

	\subsubsection{Excitation of a pair of breathers}
	
	Within the incommensurate phase,  higher values of $m_K$ correspond  to smaller solitons density [see Fig.~\ref{fig:schematics}(a)]. According to the equilibrium phase diagram, a substantial increase of lattice amplitude, $m_K^{\operatorname{init}}\ll m_K^{\text{fin}},$ can potentially trigger a transition from the incommensurate to the commensurate phase. However, this can appear differently when we consider   real-time dynamics.
	
	The only way to destroy  kinks and enter the commensurate phase is to excite the acoustic mode and  move kinks toward the boundaries, where they can disappear. However, during a quench with  $m_K^{\operatorname{init}}\ll m_K^{\text{fin}}$, the lattice amplitude increases, and each particle is trapped in a deep potential well, escaping from which it would require a high energy cost. Furthermore, during the quench of $m_K$, the solitons do not acquire a distinct momentum direction. Consequently, there are no preferred directions for their propagation. Thus, the transition to the commensurate phase remains unattainable after quenching $m_K^{\operatorname{init}}\ll m_K^{\text{fin}}$. Nonetheless, the outcome of this kind of quenches can be non-trivial, including the generation of highly non-linear types of excitations.  In particular, after increasing  $m_K$ above a certain value, a breather can be excited, see the blue region in Fig.~\ref{fig:mk}.

	The example of dynamics when a pair of breathers are excited on top of the kink is shown in Fig.~\ref{fig:breathers}(a). After increasing $m_K$, the kink in the middle of the chain remains immovable, while breathers appear on the left and right sides of the kink.  For simplicity, we mark the boundaries of the breathers in the Figure with red dashed lines. Each breather consists of a bounded kink-anti-kink pair, which propagates back and forth around a common center. This is reflected in the periodic temporal oscillation of $\phi_n$ around the breather. One can also recognize that in this particular example, the breathers slowly drift away from the center of the chain. 
	
	Qualitatively, the excitation of breathers can be understood as follows:  After increasing $m_K$, the  width of the kink, $l_m$, shrinks as $l_m\propto 1/m_K$. Consequently, immediately after the quench, particles near the center of the kink are compelled to occupy the lattice potential minima to minimize the energy. This results in the formation of a sharp kink at the center, see the blue line in Fig.~\ref{fig:breathers}(b). The displacements of the particles located immediately adjacent to the kink result in the creation of defects characterized by a negative charge. These displaced particles subsequently acquire momentum, driving them towards the system's boundaries. These anti-kink-like defects then engage with the positive-charged dislocations at the ends of the initial kink and form breathers,  illustrated by the gray regions in Fig.~\ref{fig:breathers}(b). These breathers move toward the boundaries as their initial anti-kinks constituents acquire non-zero momentum after the quench, cf.  Fig.~\ref{fig:breathers}(c). By increasing the final value of $m_K$ one can simultaneously excite a few breather modes on top of the kink.
	
	\subsection{\label{sec:TWA} Semi-classical treatment}
	
	The results discussed so far have been evaluated neglecting quantum fluctuations. However, for specific implementations, e.g., in trapped ions or cold atoms systems, quantum effects can start  playing a significant role, affecting long-time dynamics and physics close to the boundary between phases. In this case, quantum fluctuations in the model can be exponentially enhanced, inducing non-trivial dynamical responses.  
	
	The mechanism of the amplification of the quantum fluctuations in the sine-Gordon model has been shown by Starobinsky for the re-heating phase in the inflation theory in cosmology~\cite{Starobinsky}, and it can   be briefly illustrated as follows. In the low energy limit [see Eq.~\eqref{eq:ex_spectrum}], dynamics of quantum fluctuations are governed by 
	\begin{equation}\label{eq:fluctuationsdynamics}
		\ddot{\eta}_q \approx m_K^2 \cos \left(\phi^{(0)}(t)\right) \eta_q-c^2 q^2 \eta_q,
	\end{equation}
	which is valid at the early stages of the dynamics when fluctuations are small, with variance $\propto\sqrt{h_{\operatorname{eff}}}$. After the quench, the classical solution $\phi^{(0)}$ performs oscillations governed by the optical mode. Thus, the dynamics of fluctuations are similar to those of a harmonic oscillator with a periodically driven mass. Starobinsky has shown that modes with $|q| \in\left[0, m_K \sin \left| \phi^{(0)} / 2\right|\right]$ are exponentially enhanced in this case, with amplification rates $2 \Gamma_q \simeq|q| [\sin ^2\left( \phi^{(0)} / 2\right)-q^2 / m_K^2]^{1/2}$, see~\cite{Starobinsky} for more details. Amplification of these modes results in effects beyond linear response theory accompanied by the appearance of visible signatures of quantum effects in physical systems. For instance, in Ref.~\cite{PolkovnikovQBreathers}, short-lived quasi-breathers were reported in the sine-Gordon model as a manifestation of parametric resonance of quantum fluctuations in cold atoms.
	
	\bigskip
	
	\begin{figure}
		\includegraphics[width=0.99\linewidth]{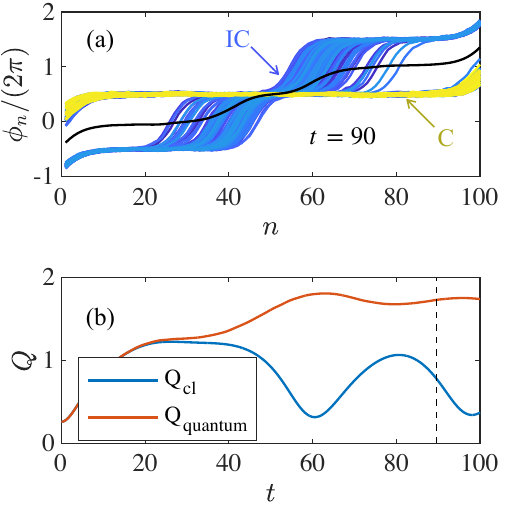}
		\caption{{Quench $m_K^{\operatorname{init}}\to m_K^f \gtrsim m_K^c>0.$ } (a) The phase  $\phi_n$ after the quench at $t=90$ computed for different noise realizations within TWA method. The resulting semi-classical configuration is plotted with the solid black line. Here $m_K^{\operatorname{init}}=0.5$, $m_K^{\operatorname{fin}}=0.29$,  $\delta=0.08$, and $h_{\operatorname{eff}}=0.07$. (b) The time dependence of the charge of the system evaluated in the classical (blue) and semi-classical limits (red). }\label{fig:semiclassics1}
	\end{figure}
	
	The dynamics of the quantum fluctuations are mostly determined by the lattice amplitude $m_K.$ For instance, if  $m_K^{\operatorname{fin}}=0$, according to Eq.~\eqref{eq:fluctuationsdynamics}, the dynamics of fluctuations and  classical phase $phi_n$ decouple, rendering the mean-field theory exact at arbitrary timescales. On the other hand, for the quench with $m_K^{\operatorname{fin}}\to 1$, the classical part $\phi_n^{(0)}$ and the quantum fluctuations $\eta$ in Eq.~\eqref{eq:fluctuationsdynamics} are strongly coupled and parametric amplification of fluctuations is possible. As a result, the larger $m_K$ is, the quicker quantum effects become visible in the dynamics.

	An example of the non-trivial manifestation of quantum effects for the quench $m_K^{\operatorname{init}}\to m_K^{\operatorname{fin}} \gtrsim m_K^c>0$  is shown in Fig.~\ref{fig:semiclassics1}. Here, we use the TWA method from Sec.~\ref{sec:preliminaries_TWA} to evaluate dynamics in the presence of quantum noise.  Panel (a)  shows the profile of the phase after the quench. Different lines result from the different initial conditions in the TWA method, and the black solid lines show the value of the phase averaged over several noise realizations. One can recognize that after this quench, some initial conditions result in commensurate configuration while others evolve towards an incommensurate state. As a result, the number of kinks averaged over several noise realizations in panel (b)  (red line) is non-integer and deviates from the value evaluated in the classical limit (blue line). The outcome of the quench is a quantum superposition of the system with and without the kink, resulting in a state without classical counterpart, as signaled by the non-integer value of the topological charge $Q$.
	
	\begin{figure}
		\includegraphics[width=0.99\linewidth]{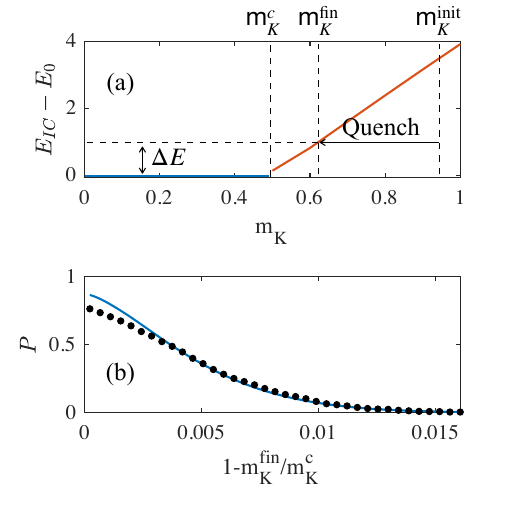}
		\caption{(a) The energy difference between the state with a kink and the ground commensurate state of the system as a function of $m_K$. The energy gap between commensurate and incommensurate configurations closes linearly with the mass of the kink. (b) The probability of nucleating a kink for a single trajectory as a function of $m_K^c-m_K^{\operatorname{fin}}$.  
		}\label{fig:semiclassics2}
	\end{figure}
	
	\bigskip
	
	To see how enhanced quantum fluctuations modify the dynamics of the Frenkel-Kontorova chain in the case above, we revisit the quench within the commensurate phase $m_K^{\operatorname{init}}\to m_K^{\operatorname{fin}} \gtrsim m_K^c>0.$ Classically, this quench keeps the system within the commensurate phase since injecting  a single kink would require an additional energy cost $\Delta E=|E\left(m_K^c\right)-E\left(m_K^{\operatorname{fin}}\right)|$, see Fig.~\ref{fig:semiclassics2}(a). Here, we plot the energy barrier the system needs to overcome to enter the incommensurate phase as a function of $m_K$. The barrier $\Delta E$ is evaluated as the energy difference between the chains with and without a kink. Adding quantum fluctuations allows the system to tunnel through this barrier and nucleate a kink at the boundary. The transition probability can be calculated with the Wentzel-Kramers-Brillouin (WKB) approximation~\cite{berry1972semiclassical}, considering the propagation of a `particle' along the $m_K$ `direction', tunneling through the barrier $\Delta E$.  This yields the formula $P=P_0 \exp (-\alpha h_{\text {eff }}^{-1} \int_{m_K^c}^{m_K^{\operatorname{fin}}} \sqrt{|E(m_K^c)-E(m_K)|} \mathrm{d} m_K)= P_0 \exp (-\alpha' |m_K^c-m_K^{\operatorname{fin}}|^{3/2})$, which we fit in good agreement with numerical results, as shown in Fig.~\ref{fig:semiclassics2}(b). This also captures that by performing a quench closer to the boundary between the two phases $m_K^c$, more final trajectories will contain a kink as they need to overcome smaller energy barriers.
	
	The effective Planck constant $h_{\operatorname{eff}}$ determines the width of the region in $m_K^c-m_K^{\operatorname{fin}}$, where a quantum superposition  takes place. Our numerical results show that $\log(P/P_0)\propto 1/\sqrt{h_{\operatorname{eff}}}$, and the effective range of $m_K^{\operatorname{fin}}$, for which nucleation of kinks occurs, is smaller than the one predicted via the WKB approximation, $\log P\propto h_{\operatorname{eff}}^{-1}$. This resulting dependence reflects that $m_K$ controls not only the dynamics of the classical part of the phase but also the coupling between the classical phase and quantum fluctuations, see Eq.~\eqref{eq:fluctuationsdynamics}. In the quenches where kinks are injected (or nucleated) into the system, the final mass is much smaller than the initial one, resulting in a suppression of quantum effects. Altogether, this provokes a more complicated dependence of $P$ on the effective Planck constant. For values of $m_K$ that are farther from the critical ones, the final state remains commensurate, and quantum noise instead provokes dephasing among trajectories at longer timescales $\propto 1/h_{\operatorname{eff}}$. The overall result is that the width of a single kink will be broadened, and the positions of particles will be more challenging to resolve.
	
	\bigskip
	
	The regime most sensitive to quantum fluctuations is the one involving breather excitations. Firstly, in the corresponding protocols, the final $m_K^{\operatorname{fin}}$ is several times larger than the initial one $m_K^{\operatorname{init}}$, thus strengthening the coupling between the classical background $\phi^{(0)}$ and quantum fluctuations $\eta$ in Eq.~\eqref{eq:fluctuationsdynamics}. Secondly, in the breather regime, $\phi^{(0)}$ oscillates, providing the mechanism for amplifying the low-energy fluctuating modes discussed above.  Thus, the effects of quantum fluctuations become visible at relatively short timescales. According to semi-classical approximation, Eq.~\eqref{eq:fluctuationsdynamics} breaks down, and different trajectories dephase. This dephasing  provokes the decay of the breather mode, making its detection more complicated. 
	
\begin{figure}
		\includegraphics[width=0.99\linewidth]{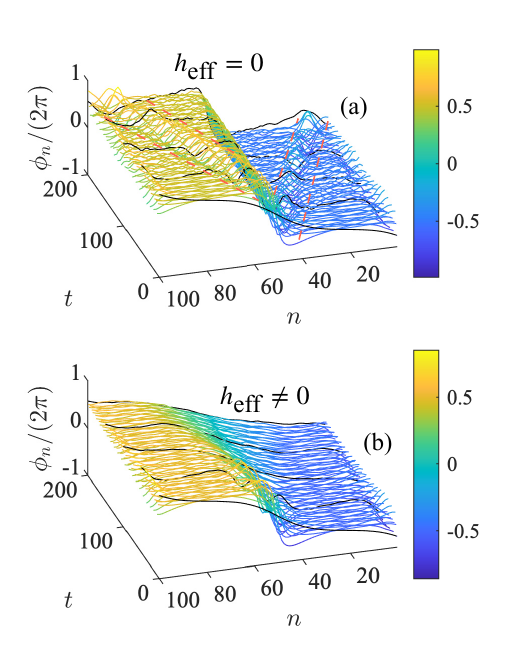}
		\caption{Evolution of the phase after the quench of $m_K$ when breather mode is excited. (a) Dynamics evaluated classically. Two breathers are excited and propagate from the center to the boundaries. (b) Semi-classical dynamics.  Due to the amplification of quantum noise on top of the oscillating background,   breathers are washed out. Here $h_{\operatorname{eff}}=0.07.$
		}\label{fig:breathers2}
	\end{figure}

	Fig.~\ref{fig:breathers2} shows the evolution of a phase where, after the quench, a pair of breathers is excited on top of the kink. For classical dynamics [see Fig.~\ref{fig:breathers2}(a)], breathers are well-localized structures that can survive for arbitrarily long times. In contrast, for semi-classical dynamics [Fig.~\ref{fig:breathers2}(b)], breathers are washed out after a few periods due to the dephasing effects. The effective lifetime of the breather mode in the presence of quantum fluctuations can be extended by working in a lower $m_K$ parameter regime and suppressing the interplay between classical and quantum parts of $\phi_n$ [cf. Eq.~\eqref{eq:fluctuationsdynamics}]. However, this regime requires a larger system size, as the size of a single kink scales as $l_m\propto 1/m_K.$
	
	\bigskip
	
	Finally, we briefly comment on the importance of the quench protocol. If, instead of the quench, we slowly vary $m_K$ -- for instance, following an adiabatic protocol -- some of the results discussed so far will be altered. The main difference  is that as the system can be adiabatically transferred between the ground states of the initial and final Hamiltonian, no oscillatory phases can be excited. This results in the absence of breathers and optical modes, thus, better system dynamics robustness against quantum fluctuations. However, the system is out of its ground state manifold after driving the commensurate-incommensurate transition, and oscillatory modes are excited, which allows quantum effects to show up at later timescales.
	
	\bigskip
	
	As shown in this section, quenching $m_K$ can lead to various outcomes, including the excitation of the optical mode, the injection of kinks, and even the generation of breathers — distinct bounded states composed of kink and anti-kink. Traditionally, breathers have been shown to emerge in kink and anti-kink scattering processes, cf. Refs.~\cite{braun2004frenkel,CARRETEROGONZALEZ2022106123,kivshar1989dynamics}. Subsequently, in the following section, we examine such processes within the framework of the FKM. Additionally, we explore another intriguing phenomenon: the scattering of a kink against a boundary. We demonstrate how the final result of this particular process is contingent on the value of the misfit parameter $\delta$.

	\section{\label{sec:Scattering}Scattering processes in the Frenkel-Kontorova model}
	
	In the integrable sine-Gordon model, the scattering of solitons is elastic~\cite{braun2004frenkel,CARRETEROGONZALEZ2022106123,kivshar1989dynamics}, while discreteness of the FKM breaks integrability and results in the distinct outputs of solitons scatterings~\cite{gomes2018false,gani2019multi,braun2004frenkel,kivshar1989dynamics,malard2013sine,braun1998nonlinear}. In this section, we probe these effects of integrability breaking on scattering processes. Firstly, we concentrate solely on discreteness effects and study kink-anti-kink scattering, comparing it to the analogous process of the sine-Gordon model. Then, we study additionally the impact of the finite size effects  by analyzing  the scattering of a kink against the boundary. We show results for zero and finite effective Planck constants in both cases. 
	
	\subsection{\label{sec:KAKScattering}Kink-anti-kink scattering} 
	
\begin{figure}
		\includegraphics[width=0.99\linewidth]{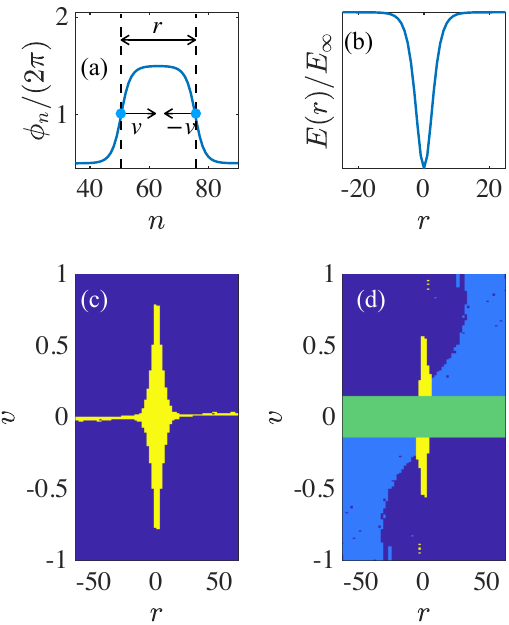}
		\caption{{Kink-anti-kink scattering.} (a) Cartoon of the scattering process.  The initial separation between solitons $r$ is the distance between their centers. The initial kink (anti-kink) velocity equals $v$ ($-v$). (b) The effective attractive potential of the kink-anti-kink pair as a function of the distance between their centers. (c) Sketch of the phase plot for kink-anti-kink scattering when $m_K=0.5\ll 1$, effectively recovering the field theory prediction. The yellow region represents the initial parameter values for which the kink-anti-kink pair evolves into the breather after   scattering, and for parameters in the dark blue region, scattering is elastic. (d) The same as in panel (c) but for $m_K=1.1$. The green region indicated velocities for which kinetic energy is smaller than $E_{PN}$, and the light blue region shows parameters for which solitons stop before the scattering.
		}\label{fig:sketchKAK}
	\end{figure}

	The sine-Gordon model  can support solutions in the form of both moving and immovable solitons, as well as bounded kink-anti-kink pairs, which are also known as breathers~\cite{etde_6020750,dauxois2006physics,braun2004frenkel}. Therefore, if we take a kink and an anti-kink that are initially separated by a distance $r$ and have initial velocities $\pm v$ [cf. Fig.~\ref{fig:sketchKAK}(a)], we can see that after the scattering, this configuration  will either evolve towards a breather or a kink-anti-kink configuration Ref.~\cite{CARRETEROGONZALEZ2022106123}.
	
	The outcome of the scattering can be predicted from   energy conservation~\cite{gani2019multi}. Fig.~\ref{fig:sketchKAK}(b) depicts the energy of the kink-anti-kink pair $E$ as a function of the distance $r$ between their centers. Here, $E_{\infty}$ denotes the energy of two infinitely separated solitons. The figure shows that at small distances, $r\lesssim l_m+l_{\delta},$   solitons experience an effective attractive potential. This suggests that if at $t=0$ the separation between the solitons is smaller than the length of a single kink, $r<l_m+l_{\delta}$,   solitons must expend the energy difference, $E_{\infty}-E_{r},$ to propagate further from the center. Therefore, if the system's kinetic energy is less than $E_{\infty}-E_{r},$   solitons remain bound within the effective attractive potential and form a breather. The yellow region in Fig.~\ref{fig:sketchKAK}(c) indicates the initial conditions for which the system evolves towards a breather state. 
	
	Conversely, if the kinetic energy exceeds $E_{\infty}-E_{r}$ [dark blue region in Fig.~\ref{fig:sketchKAK}(c)], the kink and anti-kink pass each other, with the system's final kinetic energy reduced by the value $E_{\infty}-E_{r}.$ If the initial spacing between the kink and anti-kink is larger than the length of a single kink, they scatter elastically. In this case, by moving closer to each other, they first release potential energy $E_{\infty}-E_{r}$ and accelerate. After passing each other, they pay back this energy difference from the kinetic energy, and the solitons' velocities return to their initial values.
	
	\bigskip
	
	In the discrete FKM, the picture described above remains qualitatively unchanged for small $m_K$, where the Peierls-Nabarro barrier $E_{PN}$ [Eq.~\eqref{eq:EPN}] is negligible, as shown in Fig.~\ref{fig:PN}. However, for $m_K \gtrsim 0.8$, the increase of $E_{PN}$ leads to several interesting modifications to the picture in Fig.~\ref{fig:sketchKAK}(c). The corresponding sketch of the phase plot is depicted in Fig.~\ref{fig:sketchKAK}(d). 
	
	The first major difference is that if the system's kinetic energy is less than $E_{PN}$ [green region in Fig.~\ref{fig:sketchKAK}(d)], solitons are unable to propagate through the chain (cf. Sec.~\ref{sec:PNpotential}). Such solitons remain confined at their initial locations, and the kinetic energy of the solitons is transferred,  with the single possible channel, into oscillatory optical modes. The final state is a localized kink-anti-kink pair with frozen positions of both solitons. 
	
	The second major difference occurs when solitons are widely separated at $t=0$. In this case, solitons start propagating towards each other but experience deceleration from the Peierls-Nabarro barrier (cf. Sec.~\ref{sec:PNpotential}). The kinetic energy the kink loses is proportional to the number of sites through which it propagates. Consequently, there is a critical propagation distance after   which solitons radiate a major part of their kinetic energy to the optical mode and cannot propagate further. The range of parameters $(r,v)$, corresponding  to the scenario when solitons lose all their kinetic energy before they scatter and ultimately form a localized state, are shown via light blue in Fig.~\ref{fig:sketchKAK}(d). 
	
	Finally, note that the boundary of the breather phase in the discrete model [yellow region in Fig.~\ref{fig:sketchKAK}(d)] is determined from the condition $E_{\text{kin}}-(E_{\infty}-E_{r})-E_{\operatorname{sc}} =0$, taking into account the part of the kinetic energy $E_{\operatorname{sc}}$ that system loses before the scattering due to the Peierls-Nabarro barrier.  In the continuum limit, the boundary is given simply by $E_{\text{kin}}-(E_{\infty}-E_{r})=0$.
	
	\bigskip
	
	In finite-size discrete systems, an additional modification of the kink-anti-kink scattering arises from the interaction between the solitons and the excited optical modes. To start with, consider  the propagation of one of the solitons through the discrete lattice, namely $m_K>0.8$. After the particle `jumps' from site $n_0$ to site $n_0-1$, this particle oscillates around the potential minima. Due to the nearest neighbor interaction, this oscillatory mode propagates through the chain with the speed of sound $c$. Upon reaching the boundary site $n=1$, this `wave' reflects and continues propagating in the opposite direction. Consequently, after a time interval $\Delta t_{\operatorname{opt}} =(n_0+N/2)/c$, these optical oscillations reach the center of the chain where kink-anti-kink scattering takes place. If the optical mode reaches the center of the chain faster than the corresponding kink (or anti-kink) that released this optical mode, i.e., $\Delta t_{\operatorname{opt}} < (N/2-n_0)/|v|$, one must consider the scattering of the kink, anti-kink, and the optical mode. Depending on the phase of the oscillations in an optical mode at the center of the chain during the scattering, the outcome can vary: either the optical mode provides additional energy for the solitons to overcome the effective interaction and scatter into unbounded pair, or the additional energy in the center of the chain is insufficient, and the kink-anti-kink pair evolves towards a bounded state. As a result, for small system sizes, the boundary between bounded and unbounded outcomes of the kink-anti-kink scattering may exhibit non-monotonic behavior with $r.$ 
	
	\bigskip
	
	It is also important to account for the effect of quantum fluctuations on solitons scattering. The first of such effects is the appearance of the critical region close to the separatrix between the bounded and unbounded parts in Fig.~\ref{fig:sketchKAK}(d). In the semi-classical limit, the initial conditions $(r,v)$ are defined along with uncertainties in positions and momenta on the order of $\sqrt{h_{\operatorname{eff}}}$ [cf. Eq.~\eqref{eq:TWA}]. Therefore, if the initial conditions correspond to the critical regime on the phase plot, some trajectories appear to evolve towards the breather state, while for the rest of the trajectories, the kink-anti-kink pair scatters without forming a localized state. As a result, at longer timescales, the averaged over different noise realizations phase $\phi$ evolves towards a superposition of unbounded kink-anti-kink states and breathers.
	
	An additional manifestation of quantum fluctuations appears after the scattering. This effect can be explained by the dephasing between different realizations of the dynamics that stem from different initial conditions [cf. Sec.~\ref{sec:TWA}]. Within the bounded region in Fig.~\ref{fig:sketchKAK}(d), such dephasing leads to a finite lifetime for breathers. Here, at late times the phase of breathers in each particular realization becomes significantly shifted, and after averaging over all noise realizations, the phase $\phi_n$ is constant.  Such effects have also been reported in other systems, cf. Ref.~\cite{PolkovnikovQBreathers}.  Within the unbounded region, the dephasing and variance in the initial velocities of soliton between different realizations of scattering can result in the broadening of (anti-)kink at late times. 
	
	\subsection{\label{sec:scatteringonboundary} Scattering of the kink against the boundary}
	
	\begin{figure}[]
		\includegraphics[width=0.99\linewidth]{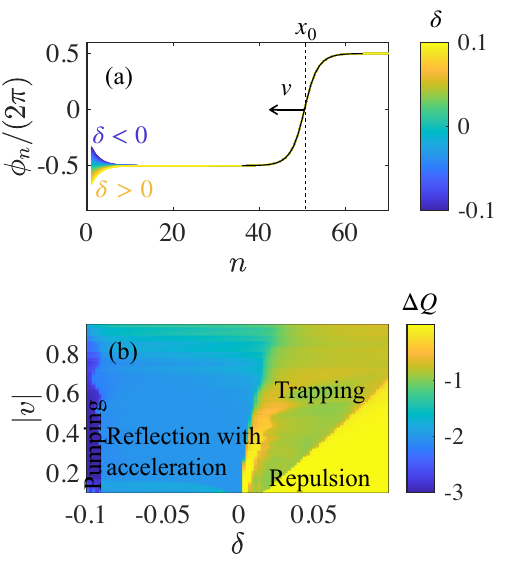}
		\caption{(a) Initial state of the system: We prepare the 
			initial configuration as a sum of the single kink with the velocity $v$ and the center at position $x_0$ (in black) and the commensurate background phase with the boundary defect. The boundary defect takes the form of the tail of kink ($\delta>0,$ yellow) or anti-kink ($\delta<0$, blue). (b) The dynamical response of the system to the scattering of the kink against the boundary defect as a function of kink velocity $v$ and misfit parameter $\delta$. Here, the final state of the system is monitored immediately after the scattering.}\label{fig:boundaryscattering}
	\end{figure}
	
	\begin{figure*}
		\includegraphics[width=0.99\linewidth]{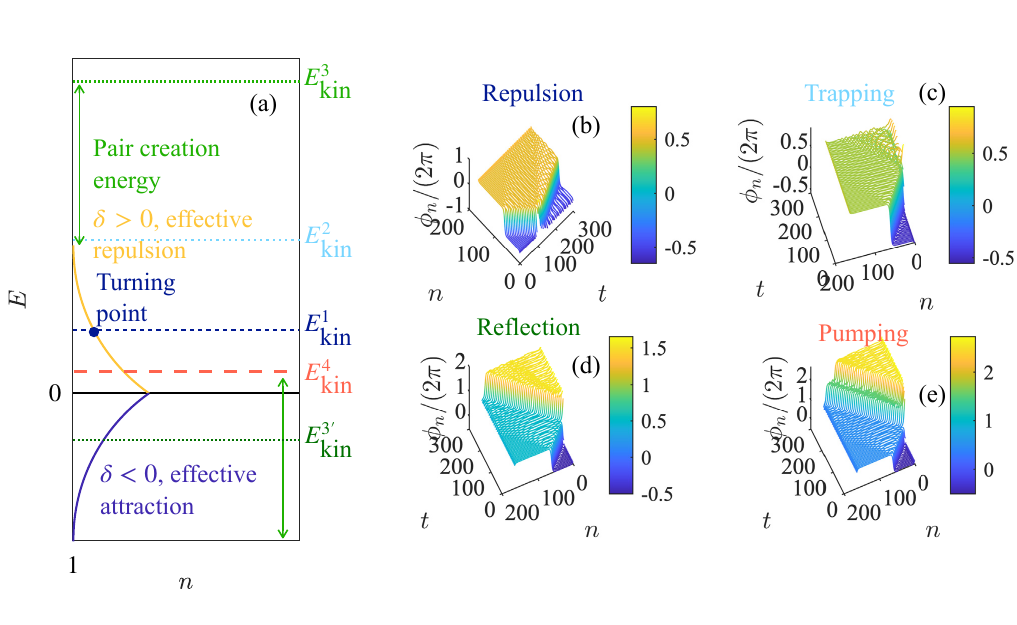}
		\caption{Scattering of a kink against the boundary. 
			(a) Sketch of the effective interaction between kink and boundary for positive (yellow) and negative (dark blue) values of misfit parameter $\delta$ experienced by the kink in close proximity to the boundary.  The instance of dynamics where kink (b)  repels from the boundary, (c) leaves the chain through  the boundary, (d) leaves the boundary and the anti-kink is injected instead, (e) leaves the boundary, and two anti-kinks are injected. }\label{fig:scattering1}
	\end{figure*}
	
	The boundary conditions of the FKM result in a unique  form of the phase  $\phi_n$ when $\delta\ne 0$. While each particle has two nearest neighbors in bulk, the edge particles interact only with the one neighbor on the left or right. Such imbalance in the interaction of edge particles [cf. Eqs.~\eqref{eq:eom}] results in the modification of the form of the  $\phi_n$ close to the boundaries in such a way that it becomes reminiscent of the tail of a single kink [for $\delta>0$, cf. yellow lines in Fig.~\ref{fig:boundaryscattering}(a)] or anti-kink [for $\delta<0$, cf.  blue lines in  Fig.~\ref{fig:boundaryscattering}(a)]. In this part of the paper, we explore  the scattering of a single kink against such unique boundary defects. This process is important because in the finite-size chain,  moving solitons reach boundaries at finite time $t~\propto N/c$ and, depending on the $\delta$, their fate therein can vary.

	To concentrate directly on the kink-boundary scattering and neglecting other kink-(anti-)kink interactions (which induce hybridization of the acoustic mode), we prepare the initial state as a combination of the commensurate background and moving kink, $\phi_n=\phi_C+\phi_{\operatorname{kink}}$. Here, $\phi_C$ represents the commensurate background phase with the boundary defect. Precisely, $\phi_C$ is a ground state solution of Eq.~\eqref{eq:eom} at a given value of $|\delta|<\delta_c$. On the other hand, $\phi_{\operatorname{kink}}$ represents a kink which we place at a specific position $x_0$ and give it a velocity of $v$ directed towards the left boundary as shown in Fig~\ref{fig:boundaryscattering}(a). This setup allows us to study the behavior of the kink as it moves towards the boundary and scatters against the boundary defect at different values of $\delta$.
	
	We fix $m_K=0.5.$ On one hand, at such amplitude of the lattice potential, the Peierls Nabarro barrier $E_{PN}$ is negligible [cf. Fig.~\ref{fig:PN}(c)], and solitons can freely propagate through the lattice without slowing down. In this way, the velocity of the kink before the scattering is equal to the initial value of $v$. On the other hand, $m_K=0.5$ is large enough to make the effect of quantum fluctuation pronounced. At this $m_K$ the effective Plank constant $h_{\operatorname{eff}}$ can be varied from $h_{\operatorname{eff}}\to0$ to some fixed value, targeting classical and semi-classical regimes of the model.
	
	\bigskip
	
	For $|\delta|<\delta_c$, there are four possible outcomes of the scattering, depending on the initial velocity of the kink $v$ and the value of the misfit parameter $\delta$. These outcomes are illustrated in Fig.~\ref{fig:boundaryscattering}(b) and include
	\begin{enumerate}
		\item Repulsion: The kink repels from the boundary and reverses its direction of propagation.
		\item Trapping: The kink leaves the chain through the boundary, and optical modes are excited.
		\item Reflection with acceleration: The kink exits the system through the boundary while an anti-kink is injected.
		\item Pumping (of two anti-kinks): Several anti-kinks are injected into the chain after the kink leaves the system.
	\end{enumerate}
	Below, we discuss each scenario separately, starting from the effects for $\delta\lesssim \delta_c$ and decreasing the misfit parameter till  $\delta\gtrsim -\delta_c.$ 
	
	\subsubsection{Repulsion}
	For $\delta>0$, the displacement of the phase on the boundary takes the form of the tail of the kink [cf. Fig.~\ref{fig:boundaryscattering}(a)]. The yellow line in Fig.~\ref{fig:scattering1}(a) shows the relationship between the system's potential energy and the kink's proximity to this kink-like boundary. When the kink approaches the boundary, it encounters an effective repulsive potential. If the initial velocity of a kink is low and the kinetic energy is smaller than the amplitude of this potential [$E^1_{\operatorname{kin}}$, blue dashed line in Fig.~\ref{fig:scattering1}(c)], the kink cannot pass the boundary. In this case, as the kink approaches the edge, it slows down, comes to a halt, and ultimately reverses its direction of propagation, see Fig.~\ref{fig:scattering1}(b). We indicate the initial parameters for which the ``Repulsion'' scenario takes place via yellow color in Fig.~\ref{fig:boundaryscattering}(b).
	
	Note that injection of kinks and excitation of the acoustic mode in Sec.~\ref{sec:sonic} at late times $t>N/c$ undergoes the scenario ``Repulsion''. Here, the injection of anti-kink or ejection of kink requires additional energy cost, which preserves the total number of solitons.
	
	\subsubsection{Trapping}
	
	If the kinetic energy of the kink is slightly higher than the effective repulsion potential [$E^2_{\operatorname{kin}}$, the light blue line in Fig.~\ref{fig:scattering1}(c)], the kink crosses the boundary and disappears. The released kinetic and potential energy of the kink is transferred into the oscillatory modes, see dynamics in Fig.~\ref{fig:scattering1}(b).  The initial parameters corresponding to this regime are shown in green in Fig.~\ref{fig:boundaryscattering}(b). 
	In this regime, quantum fluctuations may be significantly enhanced, potentially provoking dephasing effects due to the excited optical modes (cf. discussion in Sec.~\ref{sec:TWA}).

	The effective repulsion is proportional to the $\phi^2\propto\delta^2,$ and the initial kinetic energy of the kink is proportional to the $v^2.$ Thus, the critical velocity, for which the boundary repulses the kink, scales linearly with the misfit parameter, which agrees with the numerical results [see the linear boundary between the ``Trapping'' and ``Repulsion'' regions in Fig.~\ref{fig:boundaryscattering}(b)].
	
	\subsubsection{Reflection with acceleration}
	
	When the misfit parameter is negative, $\delta<0$, the phase $\phi_n$ on the boundary takes the form of the tail of an anti-kink, see Fig.~\ref{fig:boundaryscattering}(a). The potential energy of the system, in this case, decreases with the distance between the kink and the boundary [see dark blue line in Fig.~\ref{fig:scattering1}(a)]. Here, kink experiences the effective attractive potential as it comes closer to the boundary. Thus, as the kink approaches the boundary, it accelerates and exits from the chain through the edge. The energy released in this process is enough to draw another soliton into the system. Since $\delta<0$, a new anti-kink is injected into the chain. After the anti-kink enters the system, it experiences effective repulsion from the anti-kink-like boundary and accelerates toward the bulk. Overall, in this scattering process, the kink is replaced with an anti-kink, the velocity of which is higher than the initial velocity of the kink, as shown in Fig.~\ref{fig:scattering1}(d). The kinetic energy increase  comes from the effective attraction between the initial kink and anti-kink-like boundary. Note that as the effective interaction between the boundary and the kink scales like $\delta^2$, the higher the absolute value of the misfit parameter, the greater the velocity of the injected anti-kink. The initial parameters for which the ``Reflection'' scenario takes place are indicated in blue in Fig.~\ref{fig:boundaryscattering}(b).
	
	\subsubsection{Pumping}
	
	When the value of the misfit parameter is negative and close to the critical value $|\delta|\lesssim|\delta_c|$, two anti-kinks can enter the system similarly. The speed of the first anti-kink is equal to the speed of sound, and the remaining energy determines the kinetic energy of the second anti-kink. The dynamics of the system in this case are illustrated in Fig.~\ref{fig:scattering1}(e). The initial parameters for this regime are indicated in dark blue in Fig.~\ref{fig:boundaryscattering}(b).
	
	Given the finite strength of quantum fluctuations, major modifications to the picture discussed so far occur at critical regions in Fig.~\ref{fig:boundaryscattering}. Here, for parameters that correspond to the boundary between two distinct dynamical responses, the outcome of the scattering process can be the quantum superposition of these responses. 
	
	\section{\label{sec:exp}Experimental considerations}
	
	In this section, we address the experimental implementation of the FKM in the trapped ion simulator and discuss the limitations affecting the results discussed thus far. 
	
\begin{figure}
		\centering
		\includegraphics[width=0.9\linewidth]{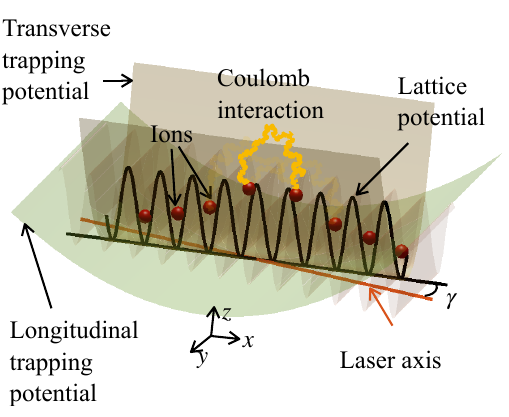}
		\caption{Sketch of the setup. Two counter-propagating laser beams (red) near $\lambda=397$nm  form lattice  potential with the periodicity $a_s=\lambda/(2\cos\gamma)$;    $\gamma$ is the angle formed between the laser beams and the $x$-axis. Ions are localized in the $yz$ plane due to the   strong trapping potential (yellow), while the weaker harmonic confinement along the $x$ (green) direction  and the mutual inter-ion Coulomb repulsion dictate the positions in the ion crystal.
		}
		\label{fig:setup_sketch}
	\end{figure}

	The trapped ion setup comprises $N$ ions confined within a trap and subjected to the lattice potential, see Fig.~\ref{fig:setup_sketch}. The ions' motion within the plane is effectively frozen due to the presence of a strong trapping potential.  Simultaneously, the particles are constrained along the $x$ axis by a weaker potential.  Thus, the kinetic energy of ions is equal to $E_k={M}\sum_{n} \dot{x}_{n}^{2}/2,$ where $M$ is the mass of a single ion, and by $x_n$ and $\dot{x}_n$ we denote the position and velocity  of $n$-th ion. The inter-particle spacing, denoted as $a_0$, characterizes the average distance between adjacent particles in the absence of the lattice potential.
	
	The mutual Coulomb repulsion can be approximated by harmonic interaction $V_{\operatorname{int}}=\sum_{n}  {K }\left({x_{n+1}-x_{n}}-a_{0}\right)^{2}/2,$ with the spring constant   $K=2 q^{2} /\left(4 \pi \varepsilon_{0} a_0^{3}\right),$ cf.~\cite{Morigi3,braun2004frenkel}. Here $q$ is the electron's charge, and $\varepsilon_0$ is the vacuum's permittivity. It is important to note that the harmonic approximation is valid only when the distance between ions, $x_{n+1}-x_n$, is much larger than  the local displacement of ions, $a_s\psi_n/(2\pi)=x_n-n a_s$. In a typical situation, higher-order corrections are suppressed by about four  orders of magnitude~\cite{marquet2003phonon}.
	
	Furthermore, ions are subjected to the periodic potential $V_{\operatorname{sub}}={\epsilon}/{2} \sum_{n}[1-\cos (2 \pi {x_{n}}/{a_{s}})],$ which is formed by the counter-propagating laser beams. Here $\epsilon$ is the lattice potential depth, and $a_s$ is  its period. The periodicity $a_s$ can be tuned by changing the relative angle $\gamma$ between the  ion axes and the propagation direction of the laser beams, $a_s=\lambda/(2\cos \gamma).$ Here $\lambda$ is the wavelength of the laser, cf. Fig.~\ref{fig:setup_sketch}.

	The total Hamiltonian reads  $H=E_k+V_{\operatorname{int}}+V_{\operatorname{sub}}$. By introducing a dimensionless  $m_{K}^{2}={(2 \pi)^{2} \epsilon}/{(2 K a_{s}^{2})}$ and $\delta={a_{0}}/{a_{s}}-1$, the dynamics of the displacement $\theta_n$ of  $n$-th ion from the $n$-th minima of the lattice potential  are described by the following Hamiltonian 
	\begin{equation}
		\begin{aligned}
			\tilde{H}=&\frac{ (2\pi)^2H}{K a_s^2}=\\
			=&\sum_n  
			\frac{1}{2} \dot{\theta}_n^2  +
			m_K^2 \left( 1+\cos\theta_n \right) +\\
			+&
			\sum_{r>0}\frac{ 1}{2  |r|^{m+2}} \left((\theta_{n+r}-\theta_{n}) -2\pi r \delta \right)^2.
		\end{aligned}
	\end{equation} 
	\noindent where $\theta_{n}=\psi_{n}+\pi$ and  $\tau=t \sqrt{K / M}$ is a  dimensionless time. In previous experiments~\cite{Ferdinand_exp,Drewsen_Dantan_2012,Vuletic_nanofriction,cetina2013one}, the inter-particle spacing was much larger than the lattice spacing, $a_0/a_s \gg 1$, such that ions are separated by a few dozens of empty lattice sites. To exclude artificial kinks corresponding to these empty sites, we can redefine variables in the following way $\phi_n\to\theta_n-2\pi n \lfloor\delta\rfloor,$ and $\delta\to \{\delta\}$, where  $\lfloor\delta\rfloor$ and $\{\delta\}$ denote integer and non-integer part of the misfit parameter respectively. Finally, by neglecting long-range interactions for $r>1$, the final Hamiltonian takes the form of~\eqref{eq:model}.
	
	\bigskip
	\begin{figure}[t]
		\centering \includegraphics[width=0.75\linewidth]{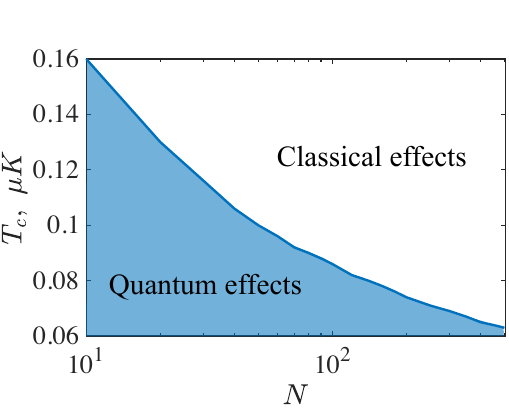}
		\caption{The dependence of the critical temperature at which thermal fluctuations dominate over quantum ones as  a function of the system size.}
		\label{fig:av_spacing}
	\end{figure}
	
	To include quantum mechanical effects into consideration, we consider quantum FKM~\cite{HU200081}, which can be obtained from~\eqref{eq:model} by  replacing classical variables with the quantum operators $x_n\to \hat{x}_n$ and $p_n\to\hat{p}_n,$ which satisfy  canonical commutation relations
	\begin{equation}\label{eq:commut}
		\begin{aligned}    
			\left[\hat{x}_n,\hat{p}_m\right]&=i \hbar\delta_{nm},\\
			\left[\hat{x}_n,\hat{x}_m\right]&=0,\\
			\left[\hat{p}_n,\hat{p}_m\right]&=0.
		\end{aligned}
	\end{equation}
	To identify the effective Planks constant, $h_{\operatorname{eff}}$, we rewrite everything in terms of the dimensionless function $\theta$ (or operator $\hat{\theta}$).  The corresponding expressions for momentum in the dimensionless form read
	\begin{equation}
		\tilde{p}_n=\dfrac{d\theta_{n}}{d\tau}, \quad \hat{\tilde{p}}_n=
		-i\dfrac{(2\pi)^2\hbar}{a_s^2\sqrt{K M}}\dfrac{d}{d\hat{\theta}_n}=-i h_{\operatorname{eff}}\dfrac{d}{d\hat{\theta}_n},
	\end{equation}
	which sets the expression for the effective Plank constant  $h_{\operatorname{eff}}={(2\pi)^2\hbar}/({a_s^2\sqrt{K M}}).$ Note that the same expression can be obtained while bringing the action to the dimensionless form, namely $\int H dt/\hbar= \int \tilde{H} d\tau /h_{\operatorname{eff}}$.

	We substitute experimental parameters from Ref.~\cite{Ferdinand_exp}, which are $a_0\in [5,15]~\mu$m, ${N}=40-100$ ions, $\epsilon \in [0,2\pi \cdot 2.7]$ {MHz}, $a_s=\lambda/(2\cos\alpha)$ where $\lambda=397$nm, $0<\alpha<22^{\circ};$ ${M}$ is the mass of $^{40}${Ca} ions, and $N\le 100$. They correspond to the following range for the dimensionless variables: $0<m_K\lesssim 1.1$, which allows us to target both ‘continuous’ and ‘discrete’ regimes from Sec.~\ref{sec:PNpotential}. The misfit parameter can be varied in the range $|\delta|<0.5$, and $h_{\operatorname{eff}}<10^{-2}$, where both genuine classical or genuine quantum mechanical effects may be observed. The  parameter regimes accessible with the trapped ion implementation of the FKM are sufficient for probing all regions in Fig.~\ref{fig:cartoon}. Furthermore, by dynamically changing the amplitude of the lattice potential, one can also potentially target the experiment regimes where solitons can dynamically injected into the chain or breathers are excited from Sec.~\ref{sec:DynamicalResponse}.

	Note that to observe the quantum superposition of states with and without the kink (see Sec.~\ref{sec:TWA}), one needs to work at a low-temperature regime. Precisely, the de Broglie wavelength $\lambda_{dB}^{\operatorname{thermal}}=[(2\pi\hbar^2)/(M k_B T)]^{1/2}$ must be smaller than the typical correlation length in the system, which is of the order of $a_s/m_K.$ Here $k_B$ is the Boltzmann constant and $T$ is the temperature. This estimation sets the upper bound of the temperature $T<10^{-9}- 10^{-7} K$. The critical temperature scales with the system size, see Fig.~\ref{fig:av_spacing}. Quantum superposition of chains with many particles requires an initialization of the system at a very low temperature. The detailed calculations are given in Appendix~\ref{sec:T_c}.
	
	Finally, in   typical experiments, the spacing between ions $a_0$ is set by competition between the  Coulomb repulsion and the external (harmonic) trapping potential [depicted in  green in Fig.~\ref{fig:setup_sketch}].  Consequently, the spacing between ions is denser at the center and wider at the borders, effectively reducing part of the crystal for which the C-IC transition can be observed (cf. Refs~\cite{Morigi_harmonic_trap,Dubin1997,Exp_long_chain}). Adjusting the setup's geometry (the periodicity $a_s$, etc.) makes it possible to separate a region with a few dozen ions where the C-IC region can be observed. At the same time, the setup with a harmonic confining potential in $x$ direction will be  unsuitable for studying a kink's scattering against the boundary: The spatial dependence of the inter-particle distance near the boundary would alter the results and induce a more complicated dependence of the kink velocity with the distance to the boundary. Alternatively, one might add optical tweezer potentials to create steep borders~\cite{tweezer1,tweezer2}, which would effectively result in the model studied in the body of the paper. In this way, the harmonic potential would be fully mitigated, and the number of ions for which the C-IC transition  takes place would largely increase.
	
	\section{Conclusions and outlook\label{sec:Conclusions}}
	
	The FKM, compared to other models that can host solitons, contains a   `chemical potential' term $\delta$, which enables the injection of kinks directly from the boundary. Without this term, the creation of solitons can be very challenging~\cite{wybo2023preparing}. On the other hand, utilizing the chemical potential enables a straightforward path toward dynamical manipulation of topological defects. 
	
	The injection of topological defects into the system changes the   spectrum of excitations and opens novel avenues toward the dynamical control of materials. The propagation of kinks can be used, for instance, to create controllable transport to reduce nano-friction when considering driven versions of the model studied here.  For instance, it has been recently proposed to use topological defects in ion crystals for enhancing heat transport~\cite{PhysRevB.108.134302}. The versatility of implementing this model in various platforms, such as  cavity QED with the Rydberg atoms~\cite{Rydberg}, cold atoms~\cite{KasperMarino,wybo2023preparing,PhysRevB.106.075102} suggest a rich array of parameters' regimes that can be targeted in experiments. One can also explore different geometries utilizing extra spatial dimensions, and study, for instance,  zigzag transition in trapped ion simulators~\cite{Giovanna_zigzag_PRA,Giovanna_zigzag_PRL,Landatrappedionzigzag,mielenz2013trapping,wybo2023preparing,PhysRevB.106.075102}.
	
	The trapped ion realization of the FKM can also be viewed as a versatile platform for modeling the physics of false vacuum decay~\cite{PhysRevD.15.2929}.  Starting in the commensurate phase, there is a range of post-quench values of $m_K^{\operatorname{fin}}\in(m_K^{cd},m_K^c]$ for which the system occupies meta-stable commensurate state, while in equilibrium these masses $m_K$ correspond to stable incommensurate phase. In this respect, after the quench, the system remains in the false vacuum (commensurate) state, while quantum fluctuations can make it decay to the stable (incommensurate) state~\cite{lagnese2021false,lagnese2023detecting}. Utilizing   Ramsay interferometry ~\cite{MorigiRamsay,Morigi_superposition} to measure this metastable state offers a tunable platform for simulating the impact on dynamics of false vacuum decay, which can have applications in high-energy physics and cosmology.
	
	\section*{Acknowledgments}
	
	We thank  R.J. Valencia-Tortora, V. Vuleti\'{c}, and A. Ziolkowska for useful discussions. 
	J. M. acknowledges E. Demler and V. Kasper for previous collaborations on related topics. 
	This project has been supported by the Deutsche Forschungsgemeinschaft (DFG, German Research Foundation) – Project-ID 429529648 – TRR 306 QuCoLiMa (``Quantum Cooperativity of Light and Matter''), and  by the Dynamics and Topology Centre funded by the State of Rhineland Palatinate and Topology Centre funded by the State of Rhineland Palatinate.
	The authors gratefully acknowledge the computing time granted on the supercomputer MOGON 2 at Johannes Gutenberg-University Mainz (hpc.uni-mainz.de). 
	
	\appendix
	
	\section{Estimation of a critical temperature via coherence resource theory\label{sec:T_c}}
	
	In this Appendix, we use the relative entropy of coherence to determine the temperature range at which quantum fluctuations dominate over thermal ones, and a quantum superposition of state with and without a kink can be dynamically generated.

	The relative entropy of coherence is a  quantifier for the resource theory of coherence which encodes the amount of superposition a given state, $\rho,$ has in a certain basis~\cite{Coherence,Shane_Coherence,huang2009introduction}. If the basis has projectors $P_d=\ket{d}\bra{d}$, then  the relative entropy of coherence is written as
	\begin{equation}
		C(\rho)=S\left(\sum_d P_d \rho P_d\right)-S\left(\rho\right),
	\end{equation}
	where $S(\rho)$ is the von Neumann  entropy of the state $\rho$.  Importantly, the state $\sum_d P_d\rho P_d$ has only diagonal entries with weights equal to the probability $P(d)=\bra{d}\rho\ket{d}$ of a measurement in the basis giving the   outcomes $d$. Thus, we find that $S\left(\sum_d P_d \rho P_d\right)$ is equal to the Shannon entropy, $H(P(d))$, of the distribution $P(d)$~\cite{bromiley2004shannon}. The relative entropy of coherence is simply the difference between the entropy of the classical distribution $P(d)$ and that of the quantum state $\rho$. If the quantum state is pure, we have  $S(\rho)=0$, and all entropy in $P(d)$ is due to superposition such that $C(\rho)=H(P(d))$; on the other hand  if the state is mixed, some of the entropy in $P(d)$ is simply due to the mixed nature of the state $\rho$, and $C(\rho)<H(P(d))$.
	
	In Sec.~\ref{sec:TWA}, we demonstrate a way of generating uncertainty in the number of solitons, $Q$, present in the system. Since $Q=(\phi_N-\phi_1)/(2\pi)$, then we should investigate the coherence in a basis such that the observables $\hat{\phi}_i$ are diagonal: $\hat{\phi}_j\ket{\{\phi_i\}}=\phi_j\ket{\{\phi_i\}}$. After the system evolves into the statistical mixture of states with and without a soliton, it will have some distribution $P(\{\phi_i\})$, for which we will want to identify how much entropy is due to superposition or a classical mixture of states: $C(\rho)=H(P(\{\phi_i\}))-S(\rho)$.
	
	While the Shannon entropy, $H\left(P\left(\{\phi_i\}\right)\right)$, is inaccessible in TWA calculations due to a sampling complexity problem, we can provide a lower bound on it using a data processing inequality~\cite{cover2012elements} applied to a statistical classifier, $f\left(\{\phi_i\}\right)\in (0,1)$ which identifies if the state has zero or one soliton. The data processing inequality states $H\left(P\left(\{\phi_i\}\right)\right)\geq H\left(P\left(f\left(\{\phi_i\}\right)\right)\right)\equiv H\left(P\left(f\right)\right)$ such that
	\begin{equation}
		C(\rho)\geq C_f\equiv H\left(P\left(f\right)\right)-S(\rho).
	\end{equation}
	When $C_f=1$, then the state is an equal superposition of a chain with and without a soliton, and can only occur when $S(\rho)=0$ (the initial state of the system is pure and the evolution over time is unitary). While $C_f$ drops below zero when initial state thermal fluctuations dominate and $S(\rho)>1.$
	
	We compute $H\left(P\left(f\right)\right)$ from $P(f=1)=1-P(f=0)$ obtained by counting the fraction of TWA trajectories with and without a soliton. To compute $S(\rho)$, we assume the initial state is in a thermal Gibbs state $\rho_0\sim e^{-\beta H}$ and undergoes strictly unitary evolution $\rho=U_t\rho_0U^\dagger_t$ where $U_t$ is the unitary generated by the Hamiltonian in Eq.~\eqref{eq:model}. The entropy of the state $\rho$ is then equivalent to the thermal entropy and can be evaluated as $S=-\sum_{q}\rho_{q}\log\rho_{q}$~\cite{huang2009introduction}, where $\rho_q={e^{-\beta H_q}}/{Z_q}$ is the reduced density matrix for mode $q$,  $\beta$ is the inverse temperature and $Z_q$ is a normalization factor.  For each mode $q$~\eqref{eq:spectrum} we have 
	\begin{equation}
		\begin{aligned}
			Z_{q}=&\operatorname{tr}\exp(-\beta H_{q})=\\
			=&\operatorname{tr}\exp\left(-\beta h_{\operatorname{eff}}\omega_{q}\left(n_{q}+\frac{1}{2}\right)\right)=\\
			=&\frac{e^{\frac{-\beta h_{\operatorname{eff}}\omega_{q}}{2}}}{1-e^{-\beta h_{\operatorname{eff}}\omega_{q}}},
		\end{aligned}
	\end{equation}
	\noindent  and after evaluating $ \rho_{q,n}=e^{-\beta h_{\operatorname{eff}}\omega_{q}n_q}\left(1-e^{-\beta h_{\operatorname{eff}}\omega_{q}}\right) $ the thermal entropy reads 
	
	\begin{equation}
		S(\rho)=\dfrac{h_{\operatorname{eff}}\omega_{q}}{k_{B}T}\langle n_{q}\rangle-\log\left(1-e^{-\beta h_{\operatorname{eff}}\omega_{q}}\right)
	\end{equation}
	\noindent with $ \langle n_{q}\rangle=(e^{\beta h_{\operatorname{eff}}\omega_{q}}-1)^{-1} .  $

	Let us consider a state that is the symmetric superposition of the commensurate and incommensurate chains, and thus $H(P(f))\approx 1$. For small temperatures, the impact of the thermal entropy is negligible, and the statistical uncertainty is solely due to superposition. Above a certain temperature scale $T_c$, $C_f$ drops below zero, indicating the state is dominated by thermal fluctuations.  For $N=10$ ions, the critical temperature is equal to $T_c=0.1\mu$K. Fig.~\ref{fig:av_spacing} illustrates the scaling of critical temperature  with the system size. Thermal entropy grows linearly with the system size; thus, at a fixed temperature, more quantum information is preserved in a system of smaller lengths. The critical temperature can be increased by enlarging the strength   of quantum fluctuations $\propto h_{\operatorname{eff}}$ at time $t=0$ or by enhancing a driving mass term for quantum fluctuations $\eta_q$ by increasing  $m_K$, see Eq.~\eqref{eq:spectrum}.
	
	\bigskip

	\bibliography{SL}
	
\end{document}